# Beyond the Local Volume. II. Population Scaleheights and Ages of Ultracool Dwarfs in Deep HST/WFC3 Parallel Fields

Christian Aganze[1], Adam J. Burgasser[1], Mathew Malkan[2], Christopher A. Theissen[1,7], Roberto A. Tejada Arevalo[3], Chih-Chun Hsu[1], Daniella C. Bardalez Gagliuffi[4], Russell E. Ryan, Jr.[5], and Benne Holwerda[6]

[1] Department of Physics, University of California San Diego, La Jolla, CA 92093, USA; caganze@ucsd.edu
[2] Department of Physics & Astronomy, University of California Los Angeles, Los Angeles, CA 90095, USA
[3] Department of Astrophysical Sciences, Princeton University, Princeton, NJ 08544, USA
[4] Department of Astrophysics, American Museum of Natural History, New York, NY 10024, USA
[5] Space Telescope Science Institute, Baltimore, MD 21218, USA
[6] Department of Physics and Astronomy, University of Louisville, Louisville, KY 40292, USA

## Abstract

Ultracool dwarfs (UCDs) represent a significant proportion of stars in the Milky Way, and deep samples of these sources have the potential to constrain the formation history and evolution of low-mass objects in the Galaxy. Until recently, spectral samples have been limited to the local volume ($d < 100$ pc). Here, we analyze a sample of 164 spectroscopically characterized UCDs identified by Aganze et al. in the Hubble Space Telescope (HST) WFC3 Infrared Spectroscopic Parallel Survey (WISPS) and 3D-HST. We model the observed luminosity function using population simulations to place constraints on scaleheights, vertical velocity dispersions, and population ages as a function of spectral type. Our star counts are consistent with a power-law mass function and constant star formation history for UCDs, with vertical scaleheights of $249^{+48}_{-61}$ pc for late-M dwarfs, $153^{+56}_{-30}$ pc for L dwarfs, and $175^{+149}_{-56}$ pc for T dwarfs. Using spatial and velocity dispersion relations, these scaleheights correspond to disk population ages of $3.6^{+0.8}_{-1.0}$ Gyr for late-M dwarfs, $2.1^{+0.9}_{-0.5}$ Gyr for L dwarfs, and $2.4^{+2.4}_{-0.8}$ Gyr for T dwarfs, which are consistent with prior simulations that predict that L-type dwarfs are on average a younger and less dispersed population. There is an additional 1–2 Gyr systematic uncertainty on these ages due to variances in age-velocity relations. We use our population simulations to predict the UCD yield in the James Webb Space Telescope PASSAGES survey, a similar and deeper survey to WISPS and 3D-HST, and find that it will produce a comparably sized UCD sample, albeit dominated by thick disk and halo sources.

*Unified Astronomy Thesaurus concepts:* Brown dwarfs (185); L dwarfs (894); T dwarfs (1679); Y dwarfs (1827); Galactic archeology (2178); Hubble Space Telescope (761); Stellar ages (1581)

*Supporting material:* machine-readable table

## 1. Introduction

The spatial and kinematic structure, chemical composition, formation, and evolutionary history of the Milky Way—an area of study known as Galactic archeology (Bland-Hawthorn & Gerhard 2016)—is probed by its stellar components (Freeman 1987; Ivezić et al. 2012). Recent advances in large-scale imaging, spectroscopic, and astrometric surveys have allowed the characterization of our galaxy at unprecedented detail. These surveys have confirmed that the stellar population of the Milky Way is grouped into four principal components: a young, metal-rich exponential thin disk; an older exponential thick disk; a diffuse, old, and metal-poor halo; and a metal-rich central bulge and bar (de Vaucouleurs & Pence 1978; Bahcall & Soneira 1981; Jurić et al. 2008; Tolstoy et al. 2009; Haywood et al. 2013). These main components contain various subpopulations and substructures, sculpted by major mergers such as Gaia-Enceladus (Belokurov et al. 2018; Helmi et al. 2018), and interactions with satellites such as Sagittarius (Price-Whelan et al. 2015; Laporte et al. 2019) and the Large and Small Magellanic Clouds (Erkal & Belokurov 2020). Other kinds of substructures can be found in abundance and kinematic patterns, including distinct chemo-kinetic populations in the halo (Helmi et al. 1999, 2018; Myeong et al. 2018; Koppelman et al. 2019; Naidu et al. 2020; Yuan et al. 2020), spatial and velocity phase-space spiral structure in the disk (Antoja et al. 2018), and numerous stellar streams (Boubert et al. 2018; Malhan et al. 2018; Shipp et al. 2018), all evidence of the complex and ongoing dynamical interactions between the Milky Way and its satellites.

Galactic spatial, kinematic, and abundance structure is typically probed through bright main-sequence FGK stars and red giants, as these stars are intrinsically bright and well distributed throughout the Galaxy. Ultracool dwarfs (UCDs), which encompass stars and brown dwarfs with $T_{\rm eff} < 3500$ K, mass $< 0.1\ M_\odot$, and spectral classes late-M, L, T, and Y (Kirkpatrick 2005), offer an alternative and complementary approach for studying the Galaxy's formation history. UCDs constitute a significant fraction of stars in the Milky Way, comprising at least 20% of all the stars in the vicinity of the Sun (Kirkpatrick et al. 2021; Reylé et al. 2021). Stellar UCDs have stable main-sequence lifetimes that far exceed the age of the universe ($\gtrsim 10^{12}$ yr), while non-fusing substellar UCDs cool continuously over time. The distinct evolution of stellar and substellar UCDs allows them to serve as *standard clocks* in

---

[7] NASA Sagan Fellow.

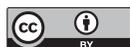

**Table 1**
Previous Deep Surveys for UCDs

| Survey Reference and Methodology | Area (deg$^2$) | Limiting Magnitude | Limiting Distance (pc) | $\log_{10}$ Effective Volume (pc$^3$)[a] | Number Detected | Population Parameters |
|---|---|---|---|---|---|---|
| Ryan et al. (2005) HST/ACS imaging | 0.038 | z (AB) < 25 | 1250 (L0) 250 (T0) | 3.1 (L0) 1.6 (T0) | 28 LT | $H = 350 \pm 50$ pc |
| Pirzkal et al. (2005) HST/ACS spectra | 0.003 | F775W (AB) < 27 | 500 (L0) 170 (L5) | 1.2 (L0) 0.1 (L5) | 18 late-M 2 L | $H = 400 \pm 100$ pc |
| Pirzkal et al. (2009) HST/ACS spectra | 0.028 | z (AB) < 25 | 1700 (M9) | 2.8 (M9) | 43 M4–M9 | $H = 370^{+60}_{-65}$ pc $H_{TD} \approx 1000$ pc $f_h \approx 0.25\%$ |
| Ryan et al. (2011) HST/WFC3 imaging | 0.064 | F125W (AB) < 25.5 F098M (AB) < 26.5 | 3000 (L0) 700 (T0) | 3.4 (L0) 2.5 (T0) | 17 MLT | $H = 290 \pm 40$ pc |
| Kakazu et al. (2010) Suprime-Cam imaging | 9.3 | z (AB) < 23.3 | 570 (L0) 120 (T0) | 4.8 (L0) 3.1 (T0) | 7 LT | $H \approx 400$ pc |
| Masters et al. (2012) HST/WFC3 spectra | 0.2 | F125W (AB) < 23 | 400 (T0) 120 (T8) | 2.8 (T0) 1.5 (T8) | 3 T | ... |
| Sorahana et al. (2019) Suprime-Cam imaging | 130 | z (AB) < 24 | 900 (L0) 238 (L8) | 6.2 (L0) 5.1 (L8) | 3,665 L | $H = 340 - 420$ pc |
| Carnero Rosell et al. (2019) Dark Energy Survey imaging | 2,400 | z (AB) < 22 | 360 (L0) 65 (T0) | 6.7 (L0) 4.8 (T0) | 11,745 LT | $H \approx 450$ pc |
| Warren et al. (2021) SDSS + UKIDSS + WISE imaging | 3,070 | J (Vega) < 17.5 | 200 (L0) | 6.3 (L0) | 1,016 L | $H = 270 \pm 6$ pc |
| Aganze et al. (2022) and this study HST/WFC3 spectra | 0.6 | F140W (AB) < 21.8 | 1170 (M7)[b] 780 (L0) 280 (T0) | 3.7 (M7)[b] 3.5 (L0)[b] 2.8 (T0)[b] | 128 late-M 26 L 10 T | $H = 249^{+48}_{-61}$ pc (late-M) $H = 153^{+56}_{-30}$ pc (L) $H = 175^{+149}_{-56}$ pc (T) |

**Notes.** This collection of surveys assumes limiting distances $\gtrsim 400$ pc to appropriately sample the vertical scaleheight of the thin disk.
[a] The effective volume of disk stars detectable in the survey, a statistic that accounts for the spatial distribution of the sample. See Section 3 for further details.
[b] Using $H = 200$ pc and one-component exponential disk model.

young stellar associations (Stauffer et al. 1998; Martín et al. 2018) and potentially in older globular clusters and the Galaxy at large (Burgasser 2004; Caiazzo et al. 2017; Gerasimov et al. 2022). Their fully convective interiors and cool, molecule-rich photospheres allow for sensitive measurement of metallicity (Rojas-Ayala et al. 2012; Veyette et al. 2017; Zhang et al. 2017), ideal for studies of chemical enrichment history. Since UCDs are more numerous and longer lived than FGK stars and red giants, they potentially offer a higher phase-space resolution of ancient spatial, kinematic, and abundance substructures, allowing us to study the early history of the Milky Way in exquisite detail.

The challenge in using UCDs for galactic archeology is their intrinsic faintness. UCD population studies based on wide-field imaging surveys have been largely limited to the nearby Solar Neighborhood ($d \lesssim 100$ pc), and have focused on measuring the local LF and velocity dispersions (Cruz et al. 2007; Metchev et al. 2008; Reyle et al. 2010; Bardalez Gagliuffi et al. 2019; Kirkpatrick et al. 2019, 2021; Burgasser et al. 2015; Hsu et al. 2021). Local studies also fail to sample the metal-poor thick disk and halo population in sufficient abundance to measure population properties; only a few dozen nearby ultracool subdwarfs have been identified to date (Burgasser et al. 2003; Lépine & Scholz 2008; Kirkpatrick et al. 2014; Zhang et al. 2019; Schneider et al. 2020).

In contrast, deep narrow-field surveys can reach more distant UCD populations, enabling measurement of disk structure and a greater proportion of halo and thick disk sources. The majority of deep surveys for UCDs (Table 1) have been undertaken with the Hubble Space Telescope (HST), as these objects often comprise a *foreground* to extragalactic surveys. Early work in this area includes measurement of M dwarf number counts in the HST Deep Field and Large Area Multi-Color Survey Groth Strip (Gould et al. 1997; Kerins 1997; Chabrier & Mera 1997). Analysis of these samples determined M dwarf thin and thick disk vertical scaleheights of ~325 pc and ~650 pc, respectively, and ruled out very low-mass stars as being an appreciable component (<1%) of Galactic halo dark matter. Ryan et al. (2005) performed one of the first deep photometric surveys of distant UCDs, identifying 28 candidate L and T dwarfs in 135 arcmin$^2$ of deep imaging data obtained with the HST Advanced Camera for Surveys (ACS) instrument, selected by their $i - z$ colors to a limiting magnitude of $z < 25$. They determined a thin disk vertical scaleheight of ~350 pc, similar to prior measurements of deep M dwarf star counts. (Ryan et al. 2011) subsequently identified 17 candidate late-M, L, and T dwarfs in 232 arcmin$^2$ of HST/Wide Field Camera 3 (WFC3) imaging of the Great Observatories Origins Deep Survey (Giavalisco et al. 2004) using optical and near-infrared color selection, and determined a thin disk vertical scaleheight for these sources of $290 \pm 40$ pc. Deep ground-based surveys have also identified samples of distant UCDs. Kakazu et al. (2010) identified seven late-L and T dwarfs in 9.3 deg$^2$ of optical and infrared imaging data from the Subaru

Suprime-Cam Hawaii Quasar and T dwarf survey to a limiting magnitude of $z < 23.3$, spectroscopically confirming several of the targets. From this small sample, Kakazu et al. (2010) inferred a thin disk vertical scaleheight of ∼400 pc for brown dwarfs. Sorahana et al. (2019) used the larger (130 deg$^2$) and deeper ($z < 24$) Hyper Suprime-Cam Subaru Strategic Program survey (Aihara et al. 2018) to photometrically identify 3,665 L dwarfs, and inferred an average thin disk vertical scaleheight of 340–420 pc. Carnero Rosell et al. (2019) used multi-band imaging data from the Dark Energy Survey (The Dark Energy Survey Collaboration 2005), combined with photometry from wide-field imaging surveys, to photometrically identify and classify 11,745 L0–T9 dwarfs to a limiting magnitude of $z \leqslant 22$, and estimated a thin disk vertical scaleheight of ∼450 pc. Recently, Warren et al. (2021) compiled a sample of 34,000 M7-L3 UCDs by searching over a large area of 3,070 deg$^2$ in the Sloan Digital Sky Survey (SDSS; York et al. 2000) and UKIRT Infrared Deep Sky Survey (UKIDSS) down to $J = 17.5$, and measured a scaleheight of ∼270 pc. These last three studies, which comprise the largest compilations of UCDs to date, use multiple colors to segregate UCDs from other background sources (Skrzypek et al. 2016).

Deep imaging surveys, typically based on optical photometry, generally probe scaleheights for only the most massive (stellar) UCDs, and are subject to significant contamination from Galactic and extragalactic populations studies. A complementary approach is to use deep spectroscopic surveys, which enable greater fidelity in both the confirmation and classification of UCDs. Pirzkal et al. (2005) identified 18 M dwarfs and two L dwarfs in the Hubble Ultra Deep Field with ACS imaging data and GRism ACS Program for Extragalactic Science (Pirzkal et al. 2004) spectroscopy. Using a single-component disk model and a halo population with a halo fraction of 0.25%, they derived a thin disk scaleheight of $400 \pm 100$ pc for their >M4 population. In a follow-up paper, Pirzkal et al. (2009) found 203 M0-M9 dwarfs in the Probing Evolution And Reionization Spectroscopically fields, based on ACS spectra and additional photometry down to $z = 25$ (AB). They found that an exponential model with two disk components was required to reproduce the observed number counts, and derived a thin disk scaleheight of $370^{+60}_{-65}$ pc for M4–M9 spectral types, a thick disk scaleheight of ∼1000 pc, and halo-to-thin disk and thick disk-to-thin disk number ratios of 0.25% and 2%, respectively, values consistent with SDSS results (Jurić et al. 2008). Pushing to lower temperatures, Masters et al. (2012) spectroscopically identified three late T dwarfs in the WFC3 Infrared Spectroscopic Parallels survey (WISP; Atek et al. 2010) based on the presence of strong $CH_4$ and $H_2O$ absorption features in 1.1–1.7 $\mu$m HST/WFC3 spectra. This small sample was sufficient to constrain the power-law index of the substellar mass function of the thin disk but not its vertical scaleheight. In Aganze et al. (2022, hereafter Paper I), we reported the discovery of 164 late-M, L, and T dwarfs in WISP and 3D-HST (Momcheva et al. 2016; Brammer et al. 2012; Skelton et al. 2014) HST/WFC3 spectroscopic data. In this study, we transform these data into the first measurement of vertical scaleheight as a function of spectral type in the UCD mass regime.

Measurement of the vertical scaleheight of UCDs as a function of spectral type or temperature is a key step toward exploring both the formation history and evolution of these objects. Vertical scaleheight is a proxy for population age (Bird et al. 2013; Mackereth et al. 2017), driven by the dynamical heating of stellar populations through encounters with Galactic structures or dispersion induced by satellite interactions (Spitzer & Schwarzschild 1953; Lacey 1984; Sellwood & Binney 2002; Hopkins et al. 2008; Ma et al. 2017). Because UCDs are a mixture of long-lived, stable stars and cooling brown dwarfs, their ages, dispersions, and scaleheights are all predicted to show complex trends with temperature depending on formation and evolutionary history (Burgasser 2004; Ryan et al. 2017). While kinematic trends can be discerned from the local sample (Zapatero Osorio et al. 2007; Blake et al. 2010; Burgasser et al. 2015; Hsu et al. 2021), variations in scaleheights require a well-characterized, large sample of UCDs out to kiloparsec distances, including low-temperature L, T, and Y dwarfs.

Section 2 describes the survey contents, including limiting magnitudes, distances, and effective volume probed. Section 3 describes the Monte Carlo population simulations used to model our star counts. Section 4 summarizes the results of our analysis, which places constraints on the scaleheights of late-M, L, and T dwarfs in the Galactic disk population. Section 5 presents predictions for the surface densities of UCDs that will be observed as part of the James Webb Space Telescope (JWST) Parallel Application of Slitless Spectroscopy to Analyze Galaxy Evolution survey (PASSAGE, JWST Cycle 1 GO-1571, PI: Malkan, Malkan et al. 2021). We summarize our main conclusions in Section 6.

## 2. Characterizing the HST/WFC3 UCD Sample

### 2.1. Observational Data

The sources and data considered in this investigation are described in Paper I. In brief, we analyzed spectral and photometric data of 164 late-M, L, and T dwarfs identified in 0.6 deg$^2$ of slitless grism spectral data obtained with HST/WFC3 in the WISP and 3D-HST surveys. The spectral data consist of low-resolution near-infrared measurements primarily spanning $1.11\,\mu m \leqslant \lambda \leqslant 1.67\,\mu m$ at an average resolution of $\lambda/\Delta\lambda \approx 130$. The photometric data include measurements in the wide-band F110W (0.8–1.4 $\mu$m), F140W (1.2-1.6 $\mu$m), and F160W (1.4–1.7 $\mu$m) filters. We also used the spectral calibration sample described in Paper I, a set of approximately 3,000 near-infrared (0.7–2.5 $\mu$m) low-resolution ($\lambda/\Delta\lambda \approx$75–120) spectra from the SpeX Prism Library (Burgasser 2014), 22 HST/WFC3 Y dwarf spectra from (Schneider et al. 2015), and 77 HST/WFC3 UCD spectra from Manjavacas et al. (2019). These data and the source properties are described in detail in Paper I.

### 2.2. Survey Limiting Magnitudes, Distances, and Effective Volumes

The number of stars present in a given field of view is equal to the number density of stars as a function of Galactic position integrated over the total volume along an observed line of sight. The volume observed is determined by the angular area of the field and the limiting distance probed, which depends on both the survey sensitivity and the intrinsic brightness of our targets. Since the latter varies as a function of spectral type in different ways between different imaging filters, it is necessary to compute these volumes as a function of spectral type and filter, in addition to considerations of varying integration times,

**Table 2**
Linear Fits to the Limiting Magnitudes of WISP Fields as a Function of G141 Exposure Time

| Filter | $c_0$ | $c_1$ | Scatter |
|---|---|---|---|
| F110W | 18.06 | 1.26 | 0.23 |
| F140W | 17.14 | 1.50 | 0.29 |
| F160W | 17.37 | 1.36 | 0.22 |

**Note.** Limiting magnitudes per filter are computed as $c_0 + c_1 \log t/1000\,s$ where $t$ is the exposure time in seconds. These relations do not account for the sensitivity correction for late spectral types discussed in Section 2.2.

photometric noise, and intrinsic variance in stellar brightness within a given spectral class, including unresolved multiplicity.

For the 3D-HST survey, Skelton et al. (2014) report the effective limiting magnitudes of each pointing based on the point-source selection criteria described in Paper I. For the WISP survey, while Atek et al. (2010) do not provide a complete list of limiting magnitudes for all pointings, they do report an average depth of F110W = 26.8 and F140W = 25.0 across the survey. Given the additional selection criteria imposed on our sample, and the differing exposure times between individual WISP and 3D-HST pointings, we chose to re-estimate the limiting magnitudes for each individual pointing and imaging filter for the WISP data. These limits were determined for a subset of pointings by fitting a Gaussian kernel density estimator (KDE; Parzen 1962) to the distribution of apparent magnitudes of point sources with a spectral $J$ signal-to-noise ratio (S/N) > 3 (see Paper I) in each pointing and filter. This fitting procedure was applied to pointings that had more than 50 qualifying point sources (dense fields) to obtain a statistically robust limit, which was estimated as the maximum of the KDE distribution. Pointings with fewer than 50 qualifying point sources (sparse fields) had faint magnitude limits estimated from linear relationships between the limiting magnitude and the logarithm of the G141 spectral exposure time from the dense pointings (see Table 2 and Figure 1). The faint magnitude limits for all fields and filters are listed in Table 3. We also adopted a bright limiting magnitude of 16 for all pointings and filters, based on the bright tails of the point-source magnitude distributions as illustrated in Figure 1.

In Paper I, we found that several of our UCD discoveries had apparent magnitudes fainter than the defined limiting magnitudes, particularly among the late-L and T dwarfs. These *deep sources* arise from the redistribution of flux within each imaging filter by strong molecular absorption from $H_2O$ and $CH_4$, in contrast to the flat, featureless spectra of most sources. The structured spectra imply that, at a given magnitude, a late-type UCD has a higher peak spectral flux density and higher $J$ S/N than the equivalent background source. In effect, we are able to probe to deeper magnitudes (and hence larger volumes) using a spectral, rather than photometric, selection criterion. To quantify this effect, we determined a limiting magnitude offset that varies with spectral type using the UCD template sample defined in Paper I. We scaled each spectrum to match the median flux in the $J$-S/N window of a standard M7 spectrum, which was assumed to have a negligible filter correction; then computed the relative magnitude in the F110W, F140W, and F160W filters. Figure 2 displays these offsets with their corresponding polynomial fits. Magnitude offsets increase with later spectral type, reaching up to ∼1.5 mag difference among late T dwarfs in the F160W band. These offsets were added to the limiting magnitudes per pointing and per spectral type in our subsequent analysis.

## 3. UCD Population Simulation

With the search volumes defined for each pointing, star counts can now be related to the luminosity function (LF), the number density of stars as a function of luminosity, brightness, temperature, or spectral type,[8] and its spatial variation (Galactic structure). These functions in turn probe UCD formation mechanisms through the mass function and birthrate; UCD thermal evolution, particularly important for brown dwarfs; and UCD dynamical evolution in the Milky Way potential. All of these factors are interdependent; hence, we follow the approach of Burgasser (2004) in simulating UCD populations through a Monte Carlo approach.

### 3.1. Simulating the UCD LF

The LF of UCDs in the immediate Solar Neighborhood ($d \lesssim 20$ pc), which largely samples the thin disk population, has been measured by several groups (Reid 2003; Cruz et al. 2003, 2007; Bochanski et al. 2010; Metchev et al. 2008; Reyle et al. 2010; Kirkpatrick et al. 2012, 2019, 2021; Burningham et al. 2013; Bardalez Gagliuffi et al. 2019). These studies reveal a UCD LF that declines from the late-M to L dwarfs as we approach the hydrogen-burning minimum mass; a minimum among the mid and late-L dwarfs, composed primarily of warm and rapidly cooling brown dwarfs; and a rise among the T and Y dwarfs as brown dwarf evolution slows. A sharp peak in the observed LF at the L dwarf/T dwarf transition reported in Kirkpatrick et al. (2019) may be a consequence of delayed evolution due to thick condensate clouds (Saumon & Marley 2008) or blended-light binaries (Burgasser 2007; see below). Integrated space densities in these samples range from $(12.6 \pm 0.6) \times 10^{-3}$ pc$^{-3}$ for M7–L5 dwarfs, $(0.5 \pm 0.3) \times 10^{-3}$ pc$^{-3}$ for L5–T6 dwarfs, and $\approx 10^{-3}$ pc$^{-3}$ for T6–Y0 dwarfs (Kirkpatrick et al. 2012, 2021; Bardalez Gagliuffi et al. 2019). The observed LFs are qualitatively and quantitatively consistent with population simulations like those described here, albeit with continued uncertainty in the form of the underlying mass function, age distribution, and role of binaries due to small samples and persistent incompleteness, even in the local volume.

In this analysis, we explicitly simulate the observed LF using assumptions of the mass function, age distribution, multiplicity, and evolutionary models. We note that Ryan & Reid (2016) use an alternative parameterized form of the local LF in their predictive study of UCDs in JWST pointings; however, this approach does not allow for evaluation of dependencies of the LF on the mass function or age distribution, nor the coupling between spatial and age distributions, which can significantly modify the observed LF in deep samples.

For our baseline LF model, we generated a sample of $10^6$ objects from a single power-law mass function parameterized by index $\alpha$ for masses nominally between $M_{low} = 0.01\,M_\odot$ and

---

[8] LFs reported in the literature are variously measured with respect to luminosity ($\frac{dN}{d\log L}$ with units of parsec$^{-3}$ dex$^{-1}$), absolute magnitude ($\frac{dN}{dM_x}$ in units of parsec$^{-3}$ mag$^{-1}$), effective temperature ($\frac{dN}{dT_{eff}}$ in units of parsec$^{-3}$ K$^{-1}$), or spectral type ($\frac{dN}{dSpT}$ in units of parsec$^{-3}$ subtype$^{-1}$), depending on the implementation. We clarify the form of the LF used with a subscript; i.e., $LF_{SpT} \equiv \frac{dN}{dSpT}$.

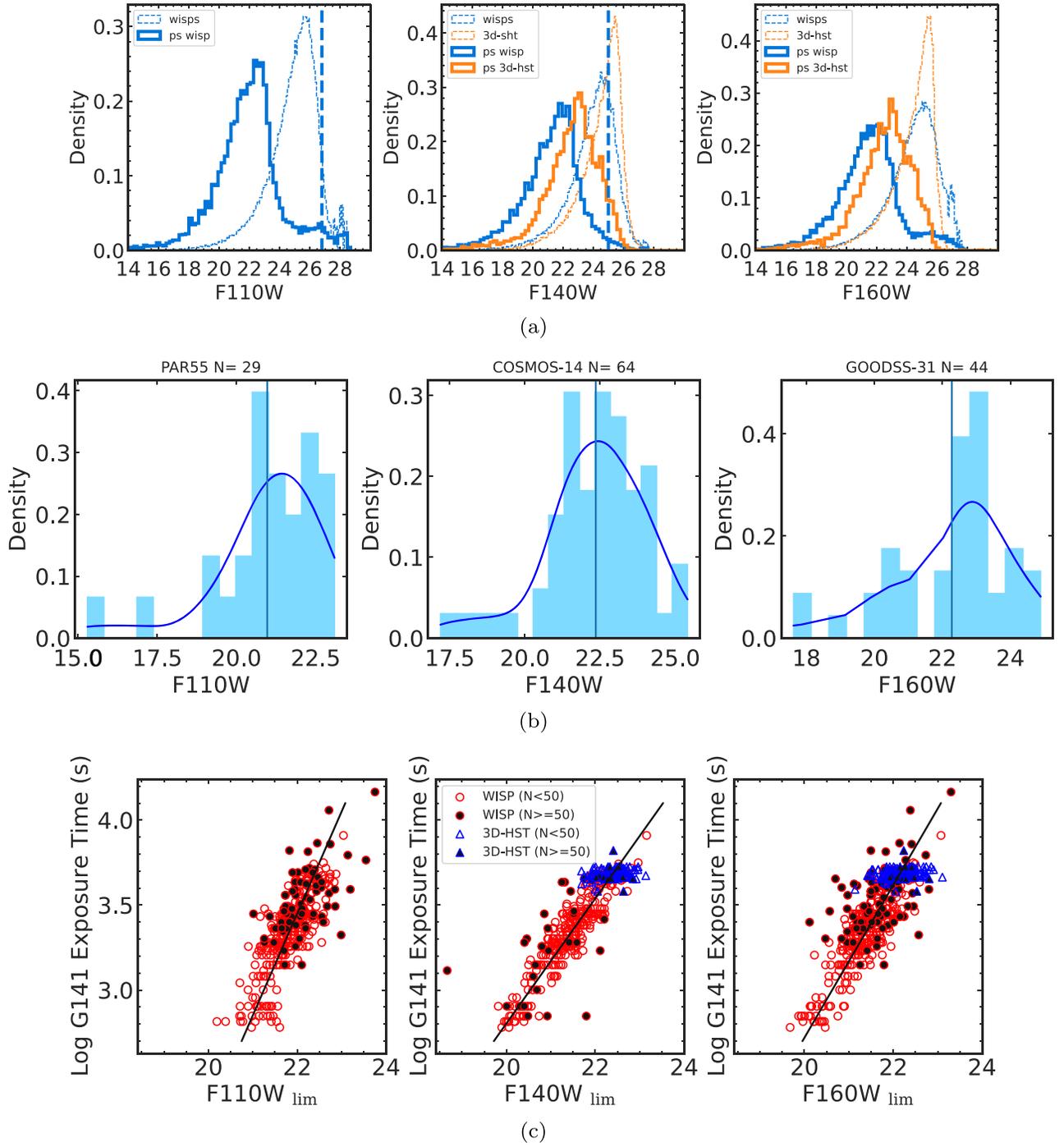

**Figure 1.** (a) Distributions of magnitudes for sources in the WISP and 3D-HST surveys across all pointings. Dotted lines show all sources, and solid lines show point sources with $J$ S/N > 3. Blue vertical dashed lines show the magnitude limits of Atek et al. (2010). (b) Estimation of the limiting magnitude for a select set of pointings. The smooth blue line shows the KDE, while the vertical black line indicates the adopted faint magnitude limit based on the KDE peak. (c) Faint magnitude limits for all pointings as a function of log spectral exposure time, separated into dense (filled symbols) and sparse (open symbols) fields. Linear fits between the dense field limits and log exposure time (black lines) were used to estimate the limiting magnitudes for the sparse fields, taking intrinsic scatter into account.

$M_{\text{high}} = 0.15\,M_\odot$:

$$P(M) = \frac{dN}{dM} \propto \left(\frac{M}{M_\odot}\right)^{-\alpha}. \quad (1)$$

We adopted $\alpha = 0.6$ based on results from (Kirkpatrick et al. 2021) for the local field population, which is also consistent with the mass functions of UCDs in young clusters (Bastian et al. 2010). Masses were drawn from this distribution by inverting the cumulative distribution function,

$$\text{CDF}(M) = \int_{M_{\text{low}}}^{M} P(m)\,dm \quad (2)$$

for $M \in [M_{\text{low}}, M_{\text{high}}]$, such that $M = \text{CDF}^{-1}(x)$ for $x \in [0, 1]$. We assigned ages using a uniform age distribution spanning 0.1–8 Gyr, which reasonably encompasses the local stellar population (Fouesneau et al. 2019).

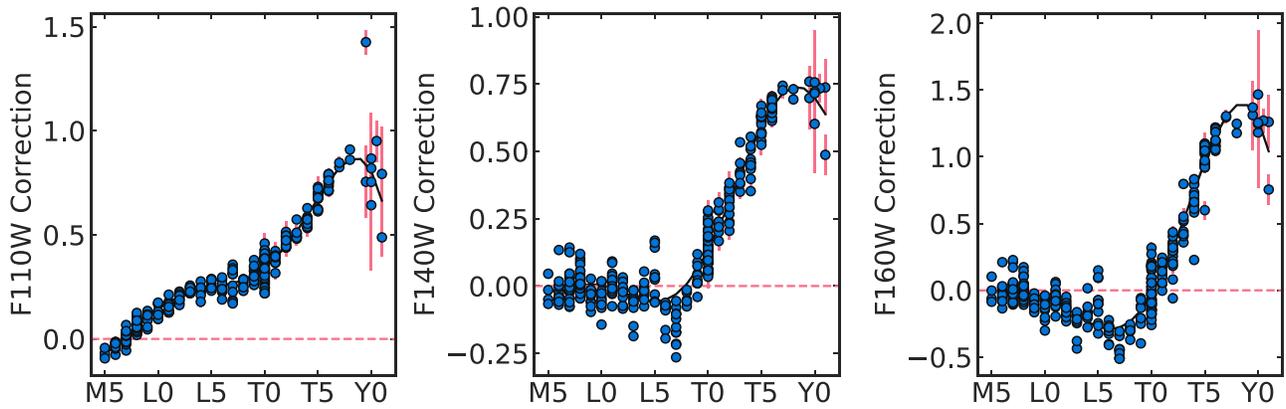

**Figure 2.** Magnitude offsets as a function of spectral type assuming a fixed $J$-S/N constraint. Black lines are fourth-order polynomial fits that take into account photometric uncertainties.

Table 3
List of Pointings Searched in This Study

| Pointing | $l$ | $b$ | G141 Exp (s) | Observation Date | F140W Exp (s) | F110W Limit (mag) | F140W Limit (mag) | F160W Limit (mag) |
|---|---|---|---|---|---|---|---|---|
| AEGIS001 | 96d26m22.7957092s | 59d29m44.83630723s | 6618.0 | 2011 May 5 | 406.0 | ⋯ | 22.4 | 22.2 |
| AEGIS002 | 96d22m11.43613852s | 59d30m03.22985808s | 5112.0 | 2011 May 3 | 812.0 | ⋯ | 22.9 | 22.4 |
| AEGIS003 | 96d26m09.58353113s | 59d40m26.09679361s | 5112.0 | 2011 Jun 13 | 812.0 | ⋯ | 22.0 | 22.5 |
| AEGIS004 | 96d29m48.36346017s | 59d36m39.08789078s | 5012.0 | 2011 Mar 16 | 812.0 | ⋯ | 22.6 | 22.0 |
| AEGIS005 | 96d29m39.13600426s | 59d39m03.51077573s | 5112.0 | 2011 Mar 16 | 812.0 | ⋯ | 22.7 | 21.7 |

**Note.** Table 3 is published in its entirety in the electronic edition of the *Astrophysical Journal*. A portion is shown here for guidance regarding its form and content.

(This table is available in its entirety in machine-readable form.)

To include the effects of multiplicity on our simulation, we assumed an overall binary fraction of 20% (Basri & Reiners 2006; Burgasser 2007; Fontanive et al. 2018), and mass ratios $q \equiv M_2/M_1$ drawn from a power-law distribution,

$$P(q) \propto q^\gamma. \quad (3)$$

We adopted $\gamma = 4$ based on the observed statistical distribution of resolved UCD binaries (Burgasser et al. 2006) with $q \in [0, 1]$. The 20% of simulated sources identified as binaries were assigned a secondary companion with mass $M_2 = qM_1$, and all binaries were assumed to be unresolved and coeval. Table 4 summarizes our population simulation parameters.

Present-day physical parameters of effective temperature ($T_{eff}$ in K), luminosity ($\log L_{bol}/L_\odot$), surface gravity ($\log g$ in centimeters per square second), and radius ($R/R_\odot$) were determined for all simulated sources and secondary components using a logarithmic interpolation of six evolutionary model grids: Burrows et al. (1997, hereafter B97), Burrows et al. (2001, hereafter B01), and Baraffe et al. (2003, hereafter B03), the hybrid cloud models of Saumon & Marley (2008, hereafter SM08), the Sonora models of Marley et al. (2018, hereafter M18), and the equilibrium chemistry ATMOS models of Phillips et al. (2020, hereafter P20). A summary of the model set assumptions and parameter limits is given in Table 5, and Figure 3 displays the grid of evolutionary used in this work. Note that some model parameter limits result in incomplete simulation samples (see discussion below). All models considered assume solar metallicity, which is appropriate given the dominance of field objects in our sample (Paper I).

Our observed sample is characterized by directly measurable quantities of spectral type and apparent magnitude; hence, it is necessary to convert simulated physical parameters into observable quantities. We used empirical relationships established from local UCD populations, specifically the spectral type to effective temperature scale of Pecaut & Mamajek (2013)[9] for objects earlier than L0; and that of Kirkpatrick et al. (2021) for L, T, and Y dwarfs. We accounted for the intrinsic scatter in these relations by random sampling.

For binary systems, we determined composite spectral types following the methodology of Burgasser (2007). We generated a sample of 104,776 binary templates from our SpeX template sample, scaling all spectra by their absolute MKO $J$ magnitudes using the $M_J$/spectral type relations from Dupuy & Liu (2012) for M6–T8 dwarfs and from Kirkpatrick et al. (2021) for T8–Y1 dwarfs. All possible pairs for which a secondary is classified as late or later than the primary were added together. Each binary template's spectral type was then determined by comparing to spectral standards, as done for our observed targets (see paper).

Figure 4 compares the mean primary, secondary, and composite classifications based on this analysis. Composite spectral types are typically ≲0.5 subtypes later than the primary type for most spectral types, with the notable exception of early T dwarf composite systems, which are up to 2.5 subtypes later. This difference is the result of the well-documented 1 $\mu$m *flux reversal* phenomenon at the L/T transition (Burgasser et al. 2002; Liu et al. 2006; Looper et al. 2008).

To scale our simulated distributions to accurately predict sample numbers, we compared our simulated effective

---
[9] https://www.pas.rochester.edu/~emamajek/EEM_dwarf_UBVIJHK_colors_Teff.txt

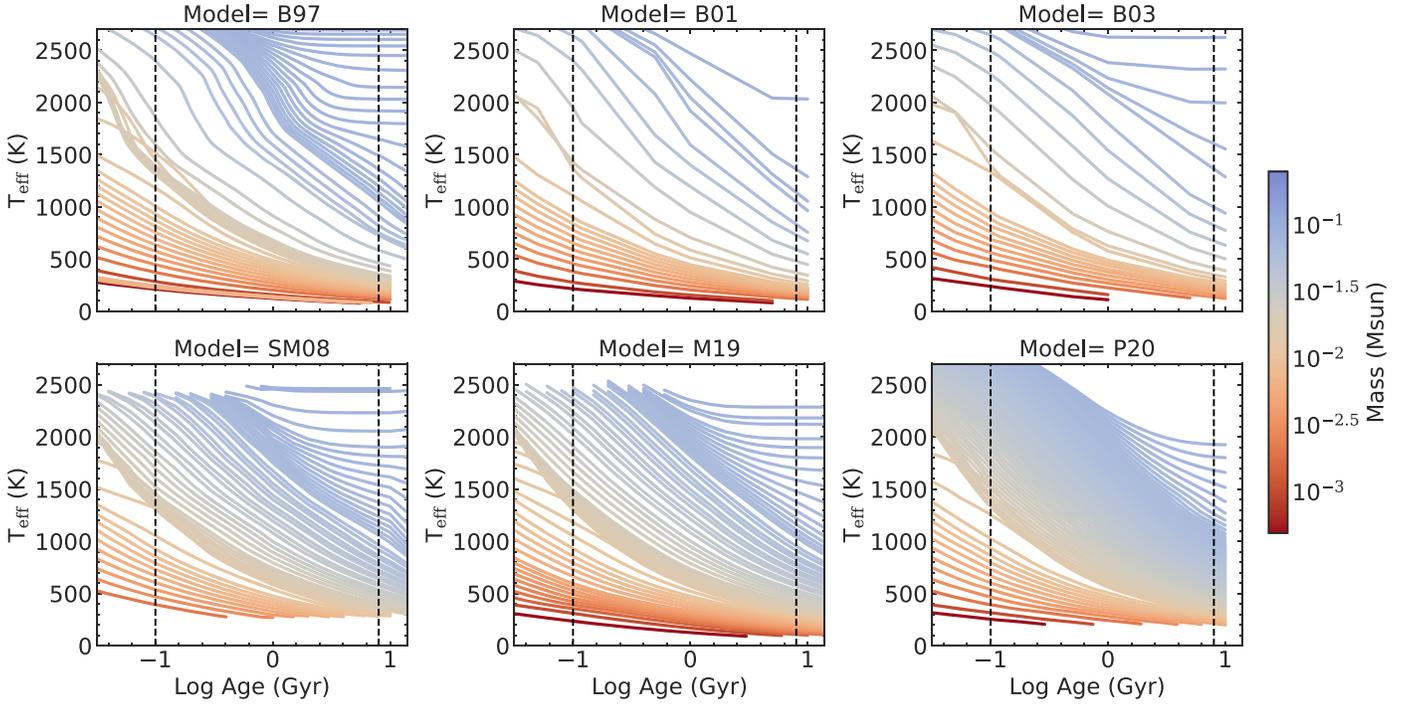

**Figure 3.** Evolutionary model grids used in this work plotted as effective temperature ($T_{eff}$) vs. age, with evolutionary isomass tracks color coded by mass. Parameter limits of these models are listed in Table 5. The dashed lines show age limits for thin disk populations from 0.1–8 Gyr.

**Table 4**
Summary of Main Parameters for Population Simulations

| Quantity | Parameterization | Parameter | Range of Quantity |
|---|---|---|---|
| Mass (M) | Power-law IMF | $\alpha = 0.6$ | $0.01\,M_\odot \leqslant M \leqslant 0.15\,M_\odot$[a] |
| Thin disk age ($\tau$) | Uniform SFH | ... | $0.1\,\mathrm{Gyr} \leqslant \tau \leqslant 8\,\mathrm{Gyr}$ |
| Binary fraction | Constant | 20% | ... |
| Mass ratio ($q = M_2/M_1$) | Power law | $\gamma = 4$ | $0 \leqslant q \leqslant 1$ |
| Evolutionary models | Grid interpolation | Various[b] | ...[c] |
| Thin disk vertical scaleheight ($H$) | Various | ... | $100\,\mathrm{pc} \leqslant H \leqslant 1000\,\mathrm{pc}$ |
| Thin disk radial scaleheight ($L$) | Constant | 2600 pc[d] | ... |
| Thick disk age ($\tau_{thick}$) | Uniform SFH | ... | $8\,\mathrm{Gyr} \leqslant \tau_{thick} \leqslant 10\,\mathrm{Gyr}$[e] |
| Thick disk vertical scaleheight ($H_{thick}$) | Constant | 900 pc[d] | ... |
| Thick disk radial scaleheight ($L_{thick}$) | Constant | 3600 pc[d] | ... |
| Thick/thin disk ratio | Constant | 12%[d], 5%[f] | ... |

**Notes.**
[a] Some evolutionary models do not cover the full mass range, see Table 5.
[b] Evolutionary models included in this analysis are from Burrows et al. (1997, 2001), Baraffe et al. (2003), Saumon & Marley (2008), Marley et al. (2018), and Phillips et al. (2020).
[c] See Table 5 for quantity ranges of individual evolutionary models.
[d] Based on Jurić et al. (2008).
[e] Based on Kilic et al. (2017).
[f] An average of the values reported by Pirzkal et al. (2009), van Vledder et al. (2016), and Hsu et al. (2021).

temperature distributions[10] $\widetilde{LF}_T$ to that reported by Kirkpatrick et al. (2021) for the local UCD population, sampling temperatures between 400 and 2000 K in $\Delta T_{eff} = 150$ K bins. We computed a scale factor ($\alpha$) that minimizes the $\chi^2$ residuals between simulated and observed LFs based on the observational uncertainties $\sigma_{LF}$:

$$\alpha = \sum_{T_i=450\,\mathrm{K}}^{T_i=2100\,\mathrm{K}} \frac{\widetilde{LF}_T \times LF_T}{\sigma_{LF}^2} \bigg/ \sum_{T_i=450\,\mathrm{K}}^{T_i=2100\,\mathrm{K}} \frac{\widetilde{LF}_T^2}{\sigma_{LF}^2} \quad (4)$$

(see, e.g., Cushing et al. 2005).

---
[10] We denote simulated distributions with the $\widetilde{LF}$ notation.

Figure 5 compares the observed and scaled simulated $LF_T$ distributions for all six evolutionary models. There is general agreement between the models, with the exception of an extra peak in the SM08 models in the 1200–1350 K bin, which is detected in the observed $LF_T$ of Kirkpatrick et al. (2021), as noted above. The absence of this feature in other evolutionary models suggests it is more likely attributed to delayed evolution at the L/T transition than multiplicity. The simulations match the observed L dwarf and late T/Y dwarf source densities, but slightly overpredict the number of mid-type T dwarfs (1000 K $\lesssim T_{eff} \lesssim$ 1500 K). We also compared the $J$-band absolute magnitude LFs ($LF_J$) of our simulations to measurements from Cruz et al. (2007) and Bardalez

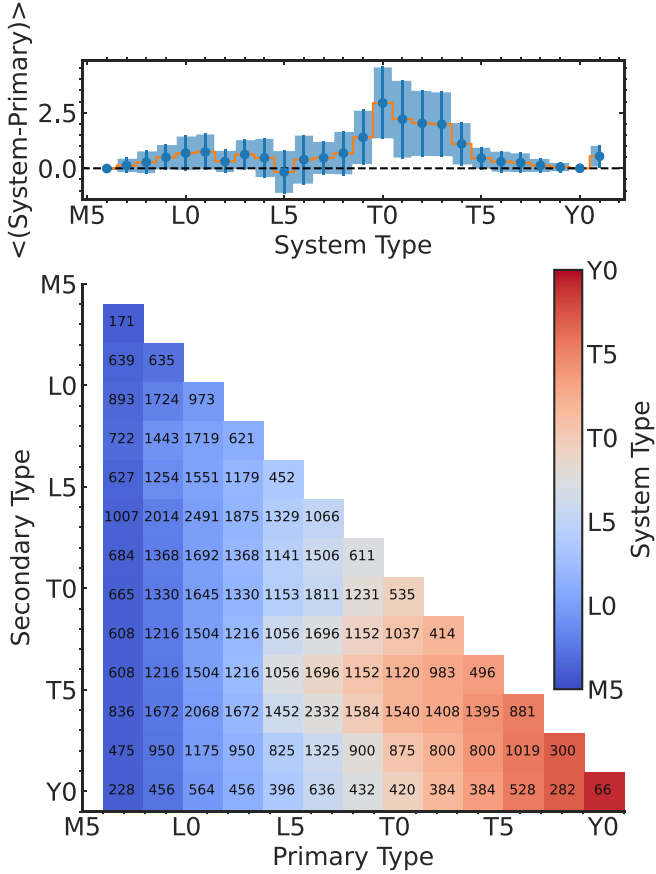

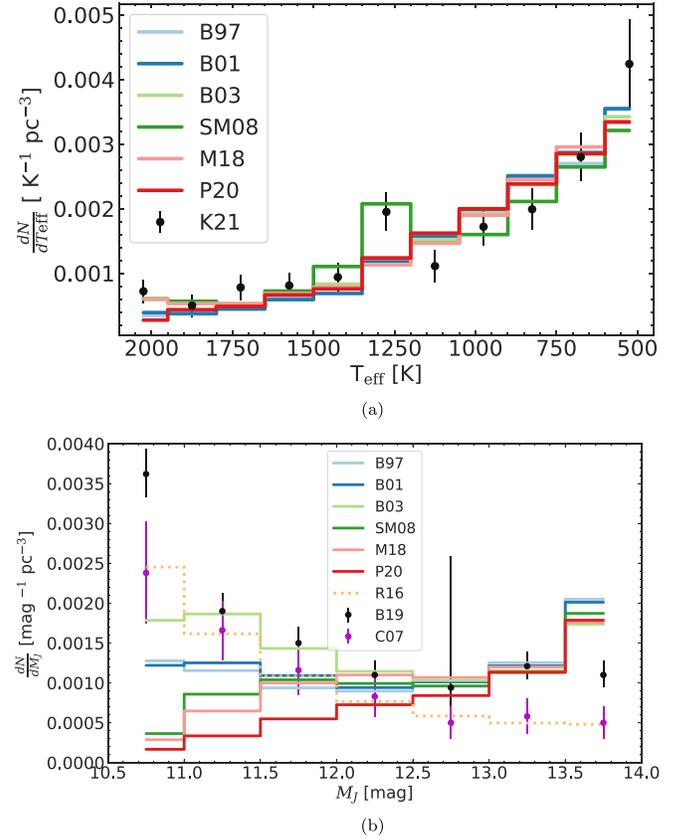

**Figure 4.** *Top panel:* mean difference between system and primary classifications as a function of system spectral type. The larger deviation of system types at T0–T2 reflects the onset of $CH_4$ absorption and the well-documented 1 $\mu$m flux reversal at the L/T transition. Uncertainties (shaded region and error bars) reflect standard deviations among the simulated binaries. *Bottom panel*: average near-infrared composite system spectral types for spectral binary templates as a function of primary and secondary classification. Note the shift toward later composite system types for mid-/late-L primaries and early/mid-T secondaries. The number of template spectra in each bin is reported. Template spectra are scaled to their absolute $M_J$ magnitudes.

**Figure 5.** (a) Comparison between our simulated LFs as a function of $T_{eff}$ ($\widetilde{LF}_T$) for different evolutionary models to the measured $LF_T$ of Kirkpatrick et al. (2021, data points with error bars). Models correspond to Burrows et al. (1997, B97), Burrows et al. (2001, B01), Baraffe et al. (2003, B03), Saumon & Marley (2008, SM08), Marley et al. (2018, M18), and Philips et al. (2020, P20). (b) Comparison between our simulated LFs as a function of absolute $J$ magnitude ($\widetilde{LF}_J$) to the observed $LF_J$s of Cruz et al. (2007, magenta points with error bars) and Bardalez Gagliuffi et al. (2019, black point with error bars). The empirical relation of Ryan & Reid (2016) is also shown as a dashed line.

Gagliuffi et al. (2019), after normalizing the simulations to the latter; and to the parameterized $LF_J$ of Ryan & Reid (2016). Here, we see significant differences between the model predictions at the brightest magnitudes, $M_J < 11$ (corresponding to spectral types <L0), largely due to parameter limits (e.g., the lack of stellar models in P20); and at the faintest magnitudes, $M_J > 13$, due to the restricted range of spectral types included in the Cruz et al. (2007) and Bardalez Gagliuffi et al. (2019) samples.

Absolute F110W, F140W, and F160W magnitudes were assigned to each simulated source using the absolute magnitude/spectral type relations developed in Paper I. We accounted for scatter in these relations by including random offsets in the assigned magnitudes following a normal distribution. For binary systems, we computed combined-light magnitudes, by adding the fluxes of the individual primary and secondary to obtain a combined magnitude in all filters.

### 3.2. Effective Volumes

The number of sources $N$(SpT) of a given spectral type detected in a field of view can be expressed as the product of the local number density of sources, $\rho_\odot$(SpT), and the effective volume of that field, $V_{eff}$(SpT). The effective volume is defined here as the density-weighted volume probed in a given radial direction based on the underlying stellar density field $\rho(\mathbf{r})$,

$$V_{eff}(SpT) = \int_V \frac{\rho(\mathbf{r})}{\rho_\odot(SpT)} dV \approx \Delta\Omega \int_{d_{min}(SpT)}^{d_{max}(SpT)} \frac{\rho(\mathbf{r})}{\rho_\odot(SpT)} r^2 dr, \quad (5)$$

where $\mathbf{r}$ is the galactocentric position vector

$$\mathbf{r} = (X, Y, Z) = (R_\odot - r\cos(b)\cos(l), \\ -r\cos(b)\sin(l), Z_\odot + r\sin(b)), \quad (6)$$

with $R_\odot = 8300$ pc and $Z_\odot = 27$ pc (Gillessen et al. 2009; Chen et al. 2001). $\Delta\Omega$ is the (assumed small) solid angle of the imaged field, equal to 4.1 arcmin$^2 = 3.47 \times 10^{-7}$ radian$^2$ for each WFC3 infrared grism image; and $r$ is the heliocentric radial distance in the direction of Galactic latitude $b$ and longitude $l$. This line integral is computed between limiting heliocentric radial distances $d_{min}(SpT) \leq r \leq d_{max}(SpT)$, which are determined from the bright and faint magnitude limits of the image ($m_{bright,faint}$; see Section 2.2) and the absolute magnitude

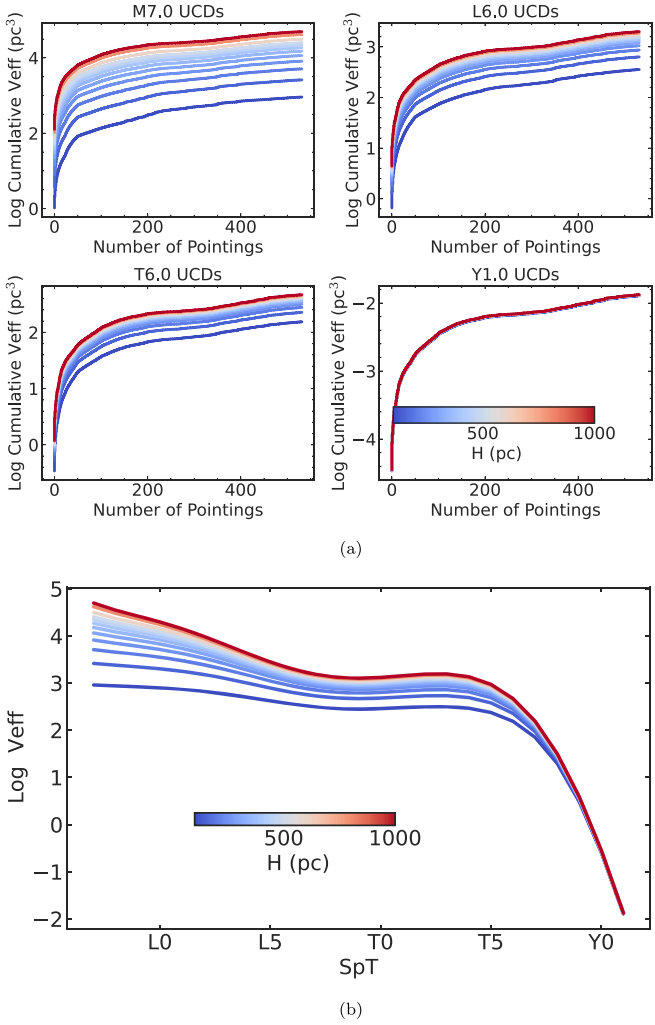

**Figure 6.** (a) The cumulative distribution of effective volumes per pointing (on a logarithmic scale) for four spectral subtypes, for various vertical scaleheights. These were constructed as a rank ordering of $V_{eff}$, such that pointings with larger effective volumes are toward the right of the distributions. The normalized forms of these distributions were used to assign pointings to our simulated sources. Note that the Y1 dwarfs show little Galactic scaleheight dependence as they are detected at close distances ($d < 10$ pc). (b) Total effective volume over all WISP and 3D-HST pointings as a function of spectral type for various vertical scaleheights.

of the source, $M(SpT)$, in a given filter:

$$\log\left(\frac{d_{\min,\max}}{10\ \text{pc}}\right)(SpT) = \frac{1}{5}(m_{\text{bright,faint}} - M(SpT)), \quad (7)$$

with $M(SpT)$ in F110W, F140W, and F160W based on the relations derived in Paper I (see above). We ignored reddening effects as all of the WISP and 3D-HST pointings are at high Galactic latitudes ($|b| > 15°$). For each pointing, we adopted the smallest outer distance limit among the imaging filters used in that pointing.

We adopted as our stellar density field a single-component, axisymmetric, exponential disk model:

$$\rho(\mathbf{r}) = \rho(R, Z) = \rho_\odot \exp\left(-\frac{R - R_\odot}{L}\right)\exp\left(-\frac{|Z - Z_\odot|}{H}\right), \quad (8)$$

where $R$, $Z$ are galactocentric cylindrical coordinates ($R^2 = X^2 + Y^2$); and $L$ and $H$ are the radial and vertical scaleheights, respectively. We adopted $L = 2600$ pc (Jurić et al. 2008), while stellar densities were computed for $H = 100$, 150, 200, 250, 300, 350, 400, 450, 500, 600, 800, and 1000 pc. Figure 6 illustrates how the total effective volume varies as a function of spectral type and vertical scaleheight.

With effective volumes computed for each pointing, spectral type, and vertical scaleheight, we assigned specific pointings to each simulated source using $V_{eff}$ as a weight factor. Specifically, we constructed cumulative distribution functions (CDFs) of the pointings by rank ordering the $V_{eff}$s for each spectral type and vertical scaleheight model. We conducted uniform draws from these CDFs to appropriately weight source locations with pointings that had larger effective volumes; i.e., those with longer exposure times (larger $d_{\max}$) and lower Galactic latitudes (smaller $|Z - Z_\odot|$).

Distances were assigned using the likelihood function

$$P(r, b, l) = \rho(\mathbf{r})r^2, \quad (9)$$

which was initially evaluated over the range $0.5d_{\min} < d < 2d_{\max}$ for each pointing and spectral type to account for sources scattered into the observed volume by photometric uncertainty or unresolved multiplicity. For each pointing and scaleheight, samples of $N = 10^5$ distances were generated by inverting the CDF associated with Equation (9) along a selected line of sight, which is equivalent to the normalized effective volume given by

$$\text{CDF}(0.5d_{\min} < r < 2d_{\max}, b, l) = \frac{\int_{0.5d_{\min}}^{r} P(r, b, l)dr}{\int_{0.5d_{\min}}^{2d_{\max}} P(r, b, l)dr}. \quad (10)$$

With these distances, individual apparent F110W, F140W, and F160W magnitudes were computed and assigned.

To account for measurement uncertainties, we fit a linear relation to the magnitude uncertainties $\sigma_m$ of all the point sources in the WISP and 3D-HST surveys as a function of filter magnitude ($m$) and log image exposure time ($t_I$),

$$\log \sigma_m = \alpha(m - m_0) - \beta \log\left(\frac{t_I}{1000\text{s}}\right) + \log \sigma_0, \quad (11)$$

where $m_0 = 19$ and $\alpha$, $\beta$, and $\sigma_0$ are fit parameters (Table 6). The magnitudes of each simulated source in all three filters are then estimated by drawing from a normal distribution with a spread equal to the appropriate magnitude uncertainty.

To appropriately model the observed spectra, we first determined a relationship between the source photometric magnitudes, grism integration times ($t_{G141}$), and $J$-S/N values (Figure 7):

$$\log J\ S/N = a\,(m - m_0) + b \log \frac{t_{G141}}{1000s} + \log J\ S/N_0, \quad (12)$$

where again $m_0 = 19$ and $a$, $b$, and $J\ S/N_0$ are the fit parameters. Here, $m$ is the simulated magnitude with uncertainty included (Equation (11)). Sources brighter than 15 mag were assigned a ceiling value of $J\ S/N = 500$, while sources fainter than 25 were assigned a floor value of $J\ S/N = 1$, matching the properties of the data. We assigned the minimum $J$-S/N value among the three imaging filters to each simulated source. The simulated apparent magnitudes and $J$ S/Ns were both used to evaluate selection effects.

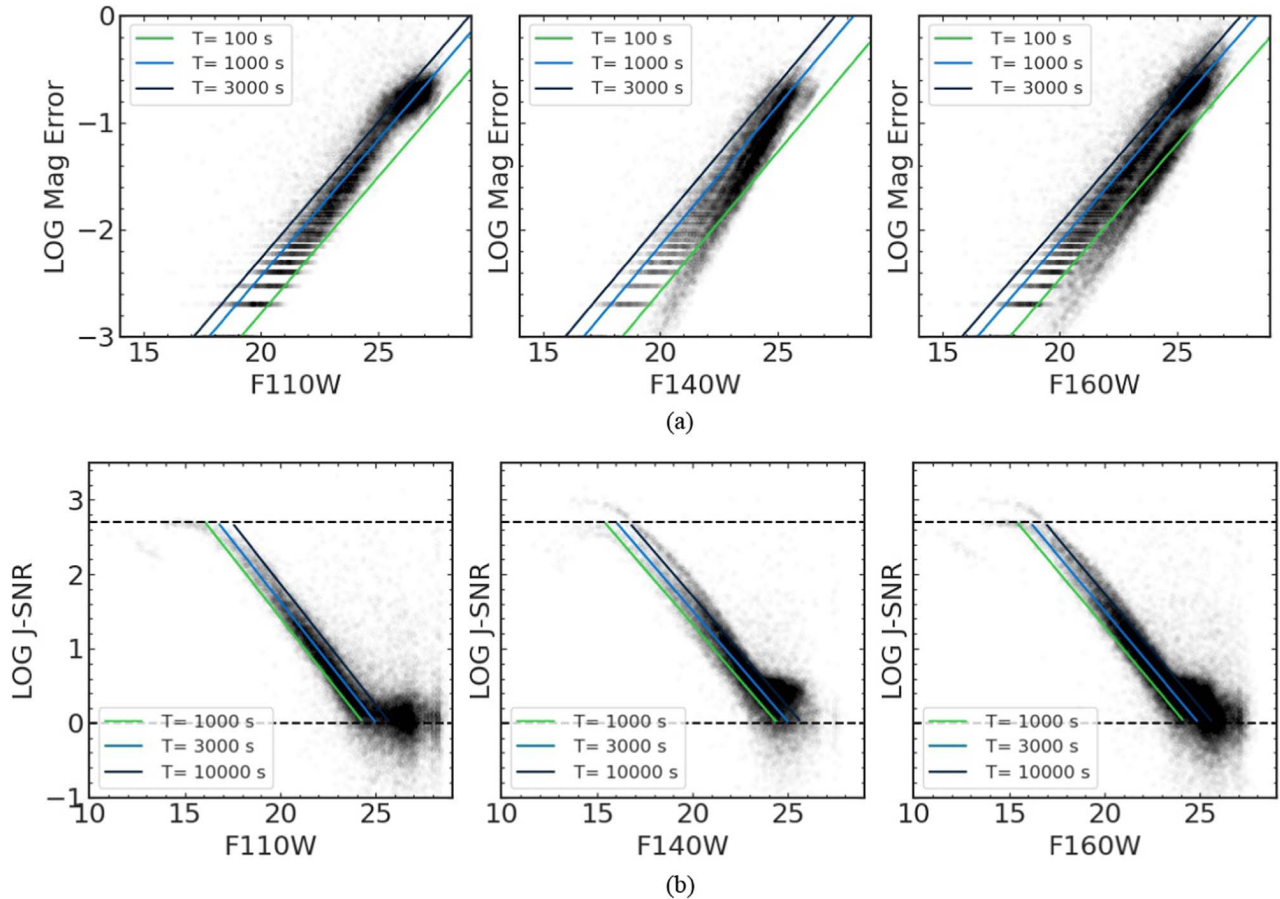

**Figure 7.** (a) Fits (straight lines) to apparent magnitude uncertainties for point sources in the WISP and 3D-HST surveys (black points) as a function of magnitude for representative exposure times of 100 s (green), 1000 s (turquoise), and 3000 s (dark blue), for F110W (left), F140W (middle), and F160W (right) imaging data. (b) Fits (straight lines) to $J$-S/N values for point-source spectra (black points) as a function of apparent magnitude for representative exposure times and the three filters (same color scheme as panel (a)). The horizontal dashed lines indicate the minimum ($J$ S/N $= 1$) and maximum ($J$ S/N $= 500$) ranges over which the fits were made.

**Table 5**
Available Range of Masses, $T_{\rm eff}$, Ages, and Cloud Treatment for Evolutionary Models

| Model | Mass | Age | $T_{\rm eff}$ | Cloud Treatment |
|---|---|---|---|---|
| B97 | $0.0005\,M_\odot \leqslant M \leqslant 0.237\,M_\odot$ | $0.0008$ Gyr $\leqslant \tau \leqslant 20$ Gy | 74 K $\leqslant \tau \leqslant$ 4363 K | Cloud-free |
| B01 | $0.0005\,M_\odot \leqslant M \leqslant 0.2\,M_\odot$ | $0.001$ Gyr $\leqslant \tau \leqslant 10$ Gyr | 82 K $\leqslant \tau \leqslant$ 4096 K | Cloud-free |
| B03 | $0.0005\,M_\odot \leqslant M \leqslant 0.1\,M_\odot$ | $0.001$ Gyr $\leqslant \tau \leqslant 10$ Gyr | 111 K $\leqslant \tau \leqslant$ 3024 K | Cloud-free |
| SM08 | $0.002\,M_\odot \leqslant M \leqslant 0.09\,M_\odot$ | $0.003$ Gyr $\leqslant \tau \leqslant 15$ Gyr | 270 K $\leqslant \tau \leqslant$ 2558 K | Hybrid |
| M18 | $0.0005\,M_\odot \leqslant M \leqslant 0.08\,M_\odot$ | $0.001$ Gyr $\leqslant \tau \leqslant 15$ Gyr | 91 K $\leqslant \tau \leqslant$ 2537 K | Cloud-free |
| P20 | $0.0005\,M_\odot \leqslant M \leqslant 0.075\,M_\odot$ | $0.001$ Gyr $\leqslant \tau \leqslant 10$ Gyr | 200 K $\leqslant \tau \leqslant$ 3075 K | Cloud-free |

### 3.3. Selection Effects

Our selection of UCDs in the WFC3 sample using indices and line and spectral template fits, and the criteria used to narrow down the sample for visual confirmation, make it likely that we rejected some fraction of UCDs present in the sample, particularly those with low S/N data. To quantify potential biases in this selection, we applied the selection criteria defined in Paper I to a sample of low-resolution UCD spectra of varying S/N to measure the recovery fraction as a function of spectral type and apparent magnitude. The simulation sample was generated from 20 of the highest S/N spectra (50 < S/N < 200) for each subtype between M7 and T9 in our template sample. For Y dwarfs, we used all templates without an S/N cut. We reduced the S/N of each spectrum by adding uncorrelated Gaussian noise, creating 100 spectra per template with $J$ S/N $\geqslant 3$.

For each of these test spectra, we computed the same indices and fit statistics, and applied the same selection criteria used to select UCDs from the WFC3 sample prior to visual confirmation, as described in Paper I. With perfect selection, we would expect to recover all spectra down to $J$ S/N $= 3$; in practice, we expect to lose some fraction of the noisiest spectra, which may depend on spectral subtype.

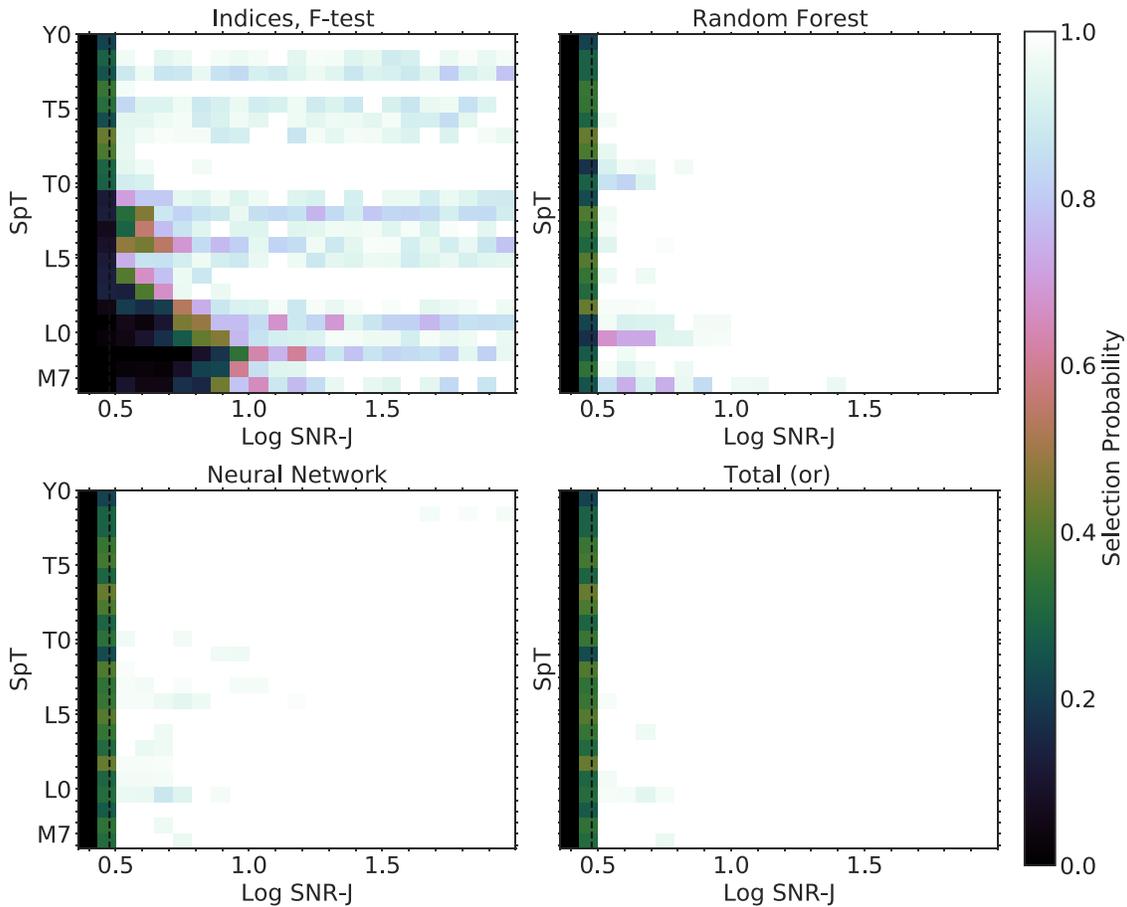

**Figure 8.** Selection probability functions for the index-based, random forest, and neural network selection methods as a function of spectral type and $J$ S/N. For the random forest and neural network models, we assumed a classification probability of 80% is required.

We defined a selection probability function ($\mathcal{S}$) as a function of $J$ S/N and spectral type as

$$\mathcal{S}(J - \mathrm{S/N}_i, \mathrm{SpT}) = \frac{N_{s,i}}{N_{\mathrm{tot},i}}, \quad (13)$$

where $N_{s,i}$ is the number of test spectra selected in a given spectral type and $J$-S/N bin $i$ (bin sizes of one subtype and two steps in S/N, respectively), and $N_{\mathrm{tot},i}$ is the total number of test spectra in that bin. *Selection* in this case means that a source is selected as any kind of UCD; e.g, a simulated M7 selected as an M9 was counted as a successful selection. For the random forest and neural network models, we used the requirement of 80% classification probability to be selected (see Paper I).

Selection probability functions for the index-based, random forest, and neural network approaches are illustrated in Figure 8. As expected, the highest S/N spectra are selected across all spectral types, but we increasingly lose objects at lower S/N values as indices become unreliable. This is particularly true for earlier spectral classes where absorption features are weaker. Sources that did not reach any of the probability cutoffs for UCDs in our selection contribute to the low-selection regions for our machine-learning methods. Note that a combination of selection criteria (as an *or* selection) reduces selection biases considerably, although this may result in much greater contamination.

With a selection probability assigned to each simulated source based on its estimated $J$ S/N and SpT, we computed the expected number of objects per spectral type in the given pointing as

$$N_{\mathrm{sim}}(\mathrm{SpT}) = \rho_\odot(\mathrm{SpT}) \cdot V_{\mathrm{eff}}(\mathrm{SpT}) \cdot \sum_i \mathcal{S}(J\ \mathrm{S/N}_i, \mathrm{SpT}), \quad (14)$$

where the sum is over all values of simulated $J$ S/N for a given SpT.

We compare these predicted number counts to the observed numbers of UCDs for each model and scaleheight in Figure 9. The observed number of UCDs in this case is not the full sample of 164 reported in Paper I, but only those sources with magnitudes brighter than the completeness limits of their respective fields in at least one of the F110W, F140W, or F160W filters, accounting for the spectral type-based correction described in Section 2.2 and displayed in Figure 2. The resulting comparison sample is composed of 98 UCDs, including 76 late-M dwarfs, 18 L dwarfs, and four T dwarfs, and has the same selection criteria imposed as our simulated sample.

### 3.4. Contamination from the Thick Disk Population

Our population simulations are designed specifically for the thin disk population, and hence will not account for thick disk or halo sources in the WFC3 survey data. To assess the contribution of the thick disk population to our sample, we performed a parallel population simulation, assuming the same

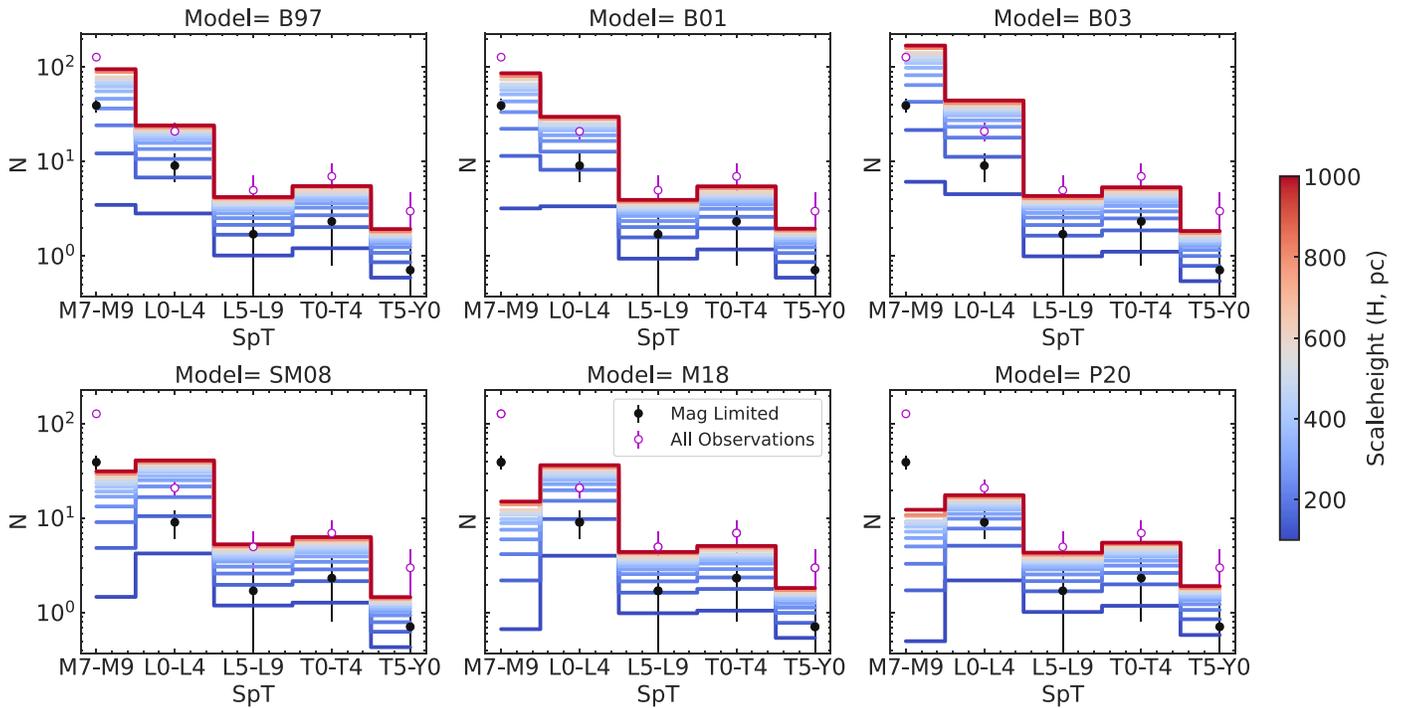

**Figure 9.** Measured UCD source counts in the combined WISP and 3D-HST fields (points with error bars) compared to predicted counts from our population simulations (histograms) for various evolutionary models (different panels) and assumed scaleheights (color scale). Number counts for our comparison sample of 98 UCDs with magnitudes brighter than the field completeness limits are shown in black points, while the full sample of 164 UCDs is shown as white points. In both cases, the estimated number of contaminating thick disk objects, assuming a thin:thick ratio of 12% has been subtracted to focus only on the thin disk predictions. Models correspond to Burrows et al. (1997, B97), Burrows et al. (2001, B01), Baraffe et al. (2003, B03), Saumon & Marley (2008, SM08), Marley et al. (2018, M18), and Phillips et al. (2020, P20).

mass range and distribution, a uniform age distribution spanning 8–10 Gyr (Kilic et al. 2017), and a fixed vertical scaleheight of $H = 900$ pc (Jurić et al. 2008). While thick disk stars are typically metal-poor and older than 10 Gyr (Gilmore et al. 1995; Haywood et al. 2013; Hawkins et al. 2015; Mackereth et al. 2017; Sharma et al. 2019), we use these solar-metallicity B03 evolutionary models and locally defined empirical calibrations as a first-order assessment. Scaling the local number density to be 12% that of the thin disk (Jurić et al. 2008), we estimate 36.7 M7–M9 dwarfs, 6.9 L0–L4 dwarfs, 0.3 L5–L9 dwarfs, and $\lesssim 1$ T and Y dwarfs in the WFC3 sample. We note that none of the UCDs identified in Paper I were matched to metal-poor subdwarf templates ([M/H] $\lesssim -1$) in the index selection analysis, although modest degrees of metal deficiency ([M/H] $\lesssim -0.5$) are likely undetectable in these data (Aganze et al. 2016). The lack of proper motion information for the vast majority of the sample also prevents us from assessing the thick disk contribution through kinematics (Bensby et al. 2003). Hsu et al. (2021) find that the fraction of intermediate thin disk/thick disk objects for UCDs ranges from 3%–8% based on the 3D kinematics of the nearby ($d < 20$ pc) UCD sample. However, these data are limited to the local volume and we expect to see a larger fraction of thick disk stars at larger distances and high galactic latitudes. (Pirzkal et al. 2009) and van Vledder et al. (2016) estimate a local thick disk fraction of M dwarfs to be 5%–7.5% (Table 1). Because of their small contribution to the sample, we chose to account for potential thick disk contamination by statistically removing these sources from our sample, rather than explicitly modeling in thick disk stars. We also ignore the contribution of halo UCDs in our sample given that the local ratio of halo-to-thin

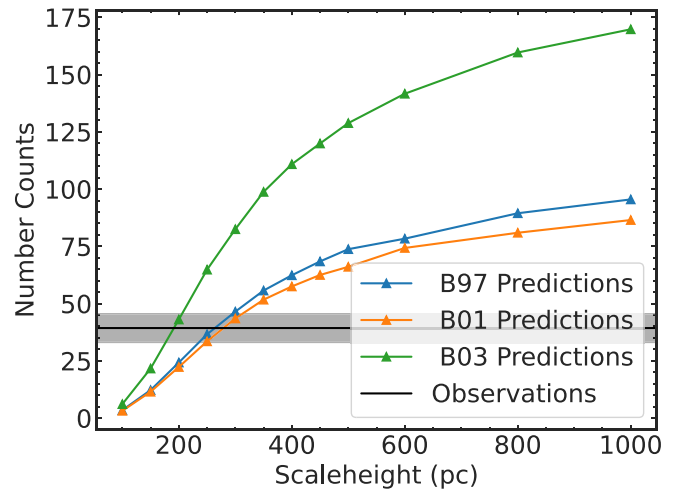

**Figure 10.** Interpolation of predicted number counts as a function of scaleheight for M7–M9 dwarfs. Predicted number counts are shown for the Burrows et al. (1997, B97; blue triangles), Burrows et al. (2001, B01; orange triangles), and Baraffe et al. (2003, B03; green triangles) evolutionary models. The black line and gray band indicate the measured number count of M7–M9 dwarfs and Poisson error.

disk stars is smaller still (0.25%–0.75%; Pirzkal et al. 2009; van Vledder et al. 2016).

## 4. Results

### 4.1. Number Counts

Figure 9 compares the predicted number of UCDs in the combined WISP and 3D-HST samples for different evolutionary models and assumed scaleheights to our observed number counts. The number counts for each subtype range is a result of competing effects between the effective volume and LF, as illustrated in Figures 3 and 4. Number counts of earlier-type UCDs (late-M and L dwarfs) are higher due to their greater limiting distances and larger volumes, while the number counts of later-type UCDs (T and Y dwarfs) are determined by their smaller volumes and the rise in the LF. The overall trend is a monotonic decrease in number counts as a function of increasing spectral type, with a slight bump for early T dwarfs.

Table 6
Fit Parameters for Simulated Magnitude Uncertainties and Spectral S/N

| Fit Quantity | Dependent Quantities | Filter | Best-fit Parameters |
| --- | --- | --- | --- |
| $\sigma_m$ | $t_I, m$ | F110W | $(\alpha, \beta, \sigma_0) = (0.25, -0.35, 0.003)$ |
| $\sigma_m$ | $t_I, m$ | F140W | $(\alpha, \beta, \sigma_0) = (0.26, -0.43, 0.007)$ |
| $\sigma_m$ | $t_I, m$ | F160W | $(\alpha, \beta, \sigma_0) = (0.25, -0.35, 0.003)$ |
| $\log J$ S/N | $t_{G141}, m$ | F110W | $(a, b, \log J\,S/N_0) = (-0.22, 0.40, 1.32)$ |
| $\log J$ S/N | $t_{G141}, m$ | F140W | $(a, b, \log J\,S/N_0) = (-0.24, 0.23, 1.33)$ |
| $\log J$ S/N | $t_{G141}, m$ | F160W | $(a, b, \log J\,S/N_0) = (-0.24, 0.39, 1.53)$ |

We see general agreement between the observed number counts and the predictions from the various evolutionary models, particularly for the L and T dwarfs. For the M7–M9 dwarfs, there are pronounced mismatches between the data and predictions from some of the models, which can be attributed to model mass and temperature parameter limits. Only the B97, B01, and B03 models encompass objects massive enough and hot enough to fully cover this spectral type range, so we restrict our analysis of the M7–M9 subtype group to these models.

### 4.2. Vertical Scaleheight

To quantify estimates of the UCD scaleheight, for each evolutionary model and spectral type subgroup, we computed a mean scaleheight by convolving the probability distribution of observed number counts with a monotonic interpolation of scaleheight as a function of predicted number counts $H(N)$, as illustrated in Figure 10. The distribution of number counts was modeled as a continuous Poisson distribution[11] assuming the observed number of thin disk sources ($N_{\text{thin}}$) to be an accurate estimate of the expected number of sources ($N$):

$$P(N) \propto \frac{N^{N_{\text{thin}}-1}}{\Gamma(N_{\text{thin}})} \exp(-N), \quad (15)$$

where $\Gamma$ is the Gamma function.

Note that the observed number of thin disk sources $N_{\text{thin}}$ is equal to the number of observed sample sources minus the predicted number of thick disk stars, $N_{\text{thin}} = N_{\text{obs}} - N_{\text{thick}}$ (Section 3.4). Drawing $10^5$ samples per model and spectral type subgroup, we inferred the distribution of scaleheights as

$$P(H) = H(P(N)). \quad (16)$$

The median values and 16% and 84% quantile uncertainties of these distributions are given in Table 7 for each model and spectral type group. We also computed median values for scaleheights per spectral type group across all models, with the exception of the M7–M9 subtype group for which we combined only the B97, B01, and B03 models.

We observe overall consistency between model predictions, with differences generally smaller than the Poisson uncertainties of the sample. One exception is the P20 model predictions for L0–L4 dwarfs, caused by the substellar upper mass limit of these models and corresponding absence of evolutionary tracks for $T_{\text{eff}} \gtrsim 2000$ K at ages $\gtrsim 1$ Gyr (Figure 3). These models were excluded in determining median values for this spectral subgroup. We also find significant differences in the inferred values of the M7–M9 dwarfs between the B97 and B01 models and the B03 models, which likely arise from differences in the evolutionary models themselves, rather than parameter limits. Since we have no a priori reason to prefer one model set over the other (however, see Section 4.4), we retain values from all three models, and our median values integrate the systemic uncertainty arising from the model differences. Finally, although the local LF predicts a bump at 1350 K (early T subtypes) specifically for SM08 models, we see a small rise in numbers for all models, so this model-dependent effect does not play a measurable role in the predicted scaleheights for our coarse subtype groupings.

Comparing our results to previously reported measurements, for the M7–M9 dwarfs we infer a vertical scaleheight of $H = 249^{+48}_{-61}$ pc, which is low but statistically consistent with values reported for late-M dwarf samples in deep HST fields by van Vledder et al. (2016, $H = 290 \pm 20$ pc for 274 M dwarfs), Ryan et al. (2005, $H = 350 \pm 50$ pc for >M6 dwarfs), Pirzkal et al. (2009, $H = 370 \pm 60$ pc for M4–M9 dwarfs), and Holwerda et al. (2014, $H = 400 \pm 100$ pc for M5–M9 dwarfs). Our results are also consistent with measurements from ground-based SDSS data reported in Bochanski et al. (2010, $H = 300 \pm 15$ pc for M0–M8 dwarfs) and SDSS-UKIDSS data reported in Warren et al. (2021, $H \approx 270$ pc for 32,942 M dwarfs). One explanation for the lower values of thin disk scaleheights for late-M dwarfs inferred in this study compared to other deep HST fields is a potential overestimation of the local fraction of thick disk objects in our sample. Additional constraints assuming a thick disk fraction of 5% are provided in Table 8. These latter values, averaging to $361^{+112}_{-125}$ pc, are more in agreement with previous deep HST field results, and indicate an inherent degeneracy between the thick disk fraction and inferred thin disk scaleheight. We will see later, however, that a small thick disk contamination is not favored in our age analysis, and for the remainder of this study, we report results that assume a thick disk fraction of 12%. Figure 11 shows our measurements of scaleheight as function of subtype and evolutionary model.

Our overall scaleheight estimates for L dwarfs ($H = 153^{+56}_{-30}$ pc) and T dwarfs ($H = 175^{+149}_{-56}$ pc) are considerably smaller than those previously inferred from deep HST ($H = 350 \pm 50$ pc; (Ryan et al. 2011) and ground-based ($H \approx 450$ pc; Carnero Rosell et al. 2019) photometric surveys

---

[11] https://docs.scipy.org/doc/scipy/reference/generated/scipy.stats.gamma.html#scipy.stats.gamma

Table 7
Scaleheights, Velocity Dispersions, and Population Ages of HST UCDs

| SpT | Quantity | B97 | B01 | B03 | SM08 | M18 | P20 | Median[a] | $N_{obs}$[b] | $N_{thick}$[c] |
|---|---|---|---|---|---|---|---|---|---|---|
| M7–M9 | $H$ (pc) | $262^{+36}_{-28}$ | $277^{+39}_{-31}$ | $191^{+16}_{-15}$ | ... | ... | ... | $249^{+48}_{-61}$ | 76 | 36.7 |
| | $\sigma_w$ (km s$^{-1}$) | $15.5^{+1.0}_{-0.9}$ | $16.0^{+1.1}_{-0.9}$ | $13.2^{+0.6}_{-0.5}$ | ... | ... | ... | $15.1^{+1.4}_{-2.0}$ | ... | ... |
| | Age (Gyr) (A09) | $3.8^{+0.6}_{-0.5}$ | $4.1^{+0.6}_{-0.5}$ | $2.7^{+0.3}_{-0.2}$ | ... | ... | ... | $3.6^{+0.8}_{-1.0}$ | ... | ... |
| | Age (Gyr) (J10) | $3.2^{+0.6}_{-0.5}$ | $3.5^{+0.7}_{-0.5}$ | $2.1^{+0.3}_{-0.2}$ | ... | ... | ... | $3.0^{+0.8}_{-1.0}$ | ... | ... |
| | Age (Gyr) (S21) | $4.9^{+0.8}_{-0.6}$ | $5.3^{+0.9}_{-0.7}$ | $3.4^{+0.4}_{-0.3}$ | ... | ... | ... | $4.6^{+1.1}_{-1.3}$ | ... | ... |
| | Median age (Gyr) (simulation) | $2.8^{+0.8}_{-0.7}$ | $3.0^{+0.8}_{-0.8}$ | $3.6^{+0.8}_{-0.7}$ | ... | ... | ... | $2.7^{+1.2}_{-2.3}$ | ... | ... |
| | K-S (A09 simulation) | **0.7** | **0.7** | **0.8** | ... | ... | ... | ... | ... | ... |
| | K-S (J10 simulation) | 0.4 | 0.4 | **0.9** | ... | ... | ... | ... | ... | ... |
| | K-S (S21 simulation) | **0.9** | **0.9** | 0.3 | ... | ... | ... | ... | ... | ... |
| L0–L4 | $H$ (pc) | $175^{+53}_{-35}$ | $156^{+39}_{-29}$ | $132^{+26}_{-20}$ | $137^{+28}_{-21}$ | $141^{+32}_{-23}$ | $222^{+118}_{-56}$ | $146^{+41}_{-27}$ | 16 | 6.9 |
| | $\sigma_w$ (km s$^{-1}$) | $12.7^{+1.8}_{-1.4}$ | $12.0^{+1.4}_{-1.2}$ | $11.0^{+1.0}_{-0.8}$ | $11.2^{+1.1}_{-0.9}$ | $11.4^{+1.2}_{-1.0}$ | $14.3^{+3.4}_{-2.0}$ | $11.6^{+1.5}_{-1.1}$ | ... | ... |
| | Age (Gyr) (A09) | $2.4^{+0.8}_{-0.5}$ | $2.1^{+0.6}_{-0.4}$ | $1.8^{+0.4}_{-0.3}$ | $1.8^{+0.4}_{-0.3}$ | $1.9^{+0.5}_{-0.4}$ | $3.2^{+2.0}_{-0.9}$ | $2.0^{+0.6}_{-0.4}$ | ... | ... |
| | Age (Gyr) (J10) | $1.8^{+0.8}_{-0.5}$ | $1.5^{+0.6}_{-0.4}$ | $1.2^{+0.4}_{-0.3}$ | $1.3^{+0.4}_{-0.3}$ | $1.3^{+0.5}_{-0.3}$ | $2.6^{+2.1}_{-0.9}$ | $1.4^{+0.6}_{-0.4}$ | ... | ... |
| | Age (Gyr) (S21) | $3.1^{+1.1}_{-0.7}$ | $2.7^{+0.8}_{-0.6}$ | $2.2^{+0.5}_{-0.4}$ | $2.3^{+0.6}_{-0.4}$ | $2.4^{+0.6}_{-0.5}$ | $4.1^{+2.6}_{-1.2}$ | $2.5^{+0.8}_{-0.5}$ | ... | ... |
| | Median age (Gyr) (simulation) | $2.3^{+1.6}_{-1.1}$ | $2.4^{+1.8}_{-1.2}$ | $2.4^{+1.7}_{-1.1}$ | $3.1^{+1.6}_{-1.5}$ | $2.4^{+1.6}_{-1.1}$ | $1.3^{+0.9}_{-0.5}$ | $2.5^{+1.7}_{-1.2}$ | ... | ... |
| | K-S (A09 simulation) | 0.2 | 0.3 | 0.4 | 0.6 | 0.4 | **0.8** | ... | ... | ... |
| | K-S (J10 simulation) | 0.3 | 0.4 | 0.6 | **0.7** | 0.6 | 0.6 | ... | ... | ... |
| | K-S (S21 simulation) | 0.4 | 0.3 | 0.3 | 0.4 | 0.2 | **0.9** | ... | ... | ... |
| L5–L9 | $H$ (pc) | $172^{+172}_{-57}$ | $176^{+199}_{-59}$ | $175^{+166}_{-59}$ | $160^{+139}_{-46}$ | $175^{+179}_{-58}$ | $174^{+190}_{-58}$ | $172^{+175}_{-56}$ | 2 | 0.3 |
| | $\sigma_w$ (km s$^{-1}$) | $12.6^{+5.2}_{-2.3}$ | $12.7^{+5.8}_{-2.4}$ | $12.7^{+5.0}_{-2.3}$ | $12.1^{+4.4}_{-1.9}$ | $12.7^{+5.4}_{-2.3}$ | $12.6^{+5.6}_{-2.3}$ | $12.6^{+5.3}_{-2.2}$ | ... | ... |
| | Age (Gyr) (A09) | $2.4^{+2.8}_{-0.9}$ | $2.4^{+3.3}_{-0.9}$ | $2.4^{+2.7}_{-0.9}$ | $2.2^{+2.2}_{-0.7}$ | $2.4^{+2.9}_{-0.9}$ | $2.4^{+3.1}_{-0.9}$ | $2.4^{+2.9}_{-0.8}$ | ... | ... |
| | Age (Gyr) (J10) | $1.8^{+3.0}_{-0.8}$ | $1.8^{+3.5}_{-0.8}$ | $1.8^{+2.8}_{-0.8}$ | $1.6^{+2.3}_{-0.6}$ | $1.8^{+3.1}_{-0.8}$ | $1.8^{+3.3}_{-0.8}$ | $1.8^{+3.0}_{-0.8}$ | ... | ... |
| | Age (Gyr) (S21) | $3.0^{+3.7}_{-1.1}$ | $3.1^{+4.4}_{-1.2}$ | $3.1^{+3.6}_{-1.2}$ | $2.8^{+3.0}_{-0.9}$ | $3.1^{+3.9}_{-1.2}$ | $3.1^{+4.1}_{-1.2}$ | $3.0^{+3.8}_{-1.1}$ | ... | ... |
| | Median age (Gyr) (simulation) | $2.4^{+1.8}_{-1.3}$ | $2.6^{+1.8}_{-1.4}$ | $2.9^{+1.8}_{-1.6}$ | $2.7^{+1.8}_{-1.5}$ | $2.6^{+1.9}_{-1.4}$ | $2.7^{+1.8}_{-1.5}$ | $2.6^{+1.8}_{-1.5}$ | ... | ... |
| | K-S (A09 simulation) | 0.2 | 0.2 | 0.1 | 0.2 | 0.2 | 0.2 | ... | ... | ... |
| | K-S (J10 simulation) | 0.2 | 0.2 | 0.2 | 0.3 | 0.2 | 0.2 | ... | ... | ... |
| | K-S (S21 simulation) | 0.3 | 0.3 | 0.2 | 0.2 | 0.2 | 0.2 | ... | ... | ... |
| T0–T4 | $H$ (pc) | $178^{+162}_{-59}$ | $182^{+183}_{-62}$ | $185^{+180}_{-64}$ | $173^{+150}_{-55}$ | $191^{+180}_{-69}$ | $180^{+165}_{-61}$ | $181^{+169}_{-62}$ | 3 | 0.7 |
| | $\sigma_w$ (km s$^{-1}$) | $12.8^{+4.9}_{-2.3}$ | $12.9^{+5.4}_{-2.4}$ | $13.0^{+5.3}_{-2.5}$ | $12.6^{+4.6}_{-2.2}$ | $13.2^{+5.2}_{-2.7}$ | $12.9^{+4.9}_{-2.4}$ | $12.9^{+5.0}_{-2.4}$ | ... | ... |
| | Age (Gyr) (A09) | $2.5^{+2.6}_{-0.9}$ | $2.5^{+3.0}_{-1.0}$ | $2.6^{+3.0}_{-1.0}$ | $2.4^{+2.4}_{-0.8}$ | $2.7^{+3.0}_{-1.1}$ | $2.5^{+2.7}_{-0.9}$ | $2.5^{+2.8}_{-0.9}$ | ... | ... |
| | Age (Gyr) (J10) | $1.9^{+2.8}_{-0.8}$ | $1.9^{+3.2}_{-0.9}$ | $2.0^{+3.2}_{-0.9}$ | $1.8^{+2.6}_{-0.8}$ | $2.1^{+3.2}_{-1.0}$ | $1.9^{+2.8}_{-0.9}$ | $1.9^{+2.9}_{-0.9}$ | ... | ... |
| | Age (Gyr) (S21) | $3.1^{+3.5}_{-1.2}$ | $3.2^{+4.0}_{-1.3}$ | $3.3^{+3.9}_{-1.3}$ | $3.0^{+3.2}_{-1.1}$ | $3.4^{+4.0}_{-1.4}$ | $3.2^{+3.6}_{-1.2}$ | $3.2^{+3.7}_{-1.2}$ | ... | ... |
| | Median age (Gyr) (simulation) | $3.2^{+1.7}_{-1.6}$ | $3.2^{+1.8}_{-1.6}$ | $3.4^{+1.8}_{-1.7}$ | $3.2^{+1.8}_{-1.6}$ | $3.2^{+1.8}_{-1.6}$ | $3.5^{+1.8}_{-1.7}$ | $3.3^{+1.8}_{-1.6}$ | ... | ... |
| | K-S (A09 simulation) | 0.2 | 0.2 | 0.2 | 0.2 | 0.1 | 0.2 | ... | ... | ... |
| | K-S (J10 simulation) | 0.3 | 0.3 | 0.3 | 0.4 | 0.3 | 0.4 | ... | ... | ... |
| | K-S (S21 simulation) | 0.1 | 0.1 | 0.1 | 0.1 | 0.2 | 0.1 | ... | ... | ... |
| T5–Y0 | $H$ (pc) | $183^{+241}_{-66}$ | $182^{+227}_{-65}$ | $188^{+269}_{-69}$ | $196^{+229}_{-75}$ | $188^{+230}_{-69}$ | $183^{+234}_{-65}$ | $187^{+237}_{-68}$ | 1 | 0.3 |
| | $\sigma_w$ (km s$^{-1}$) | $13.0^{+6.8}_{-2.6}$ | $13.0^{+6.4}_{-2.6}$ | $13.1^{+7.4}_{-2.7}$ | $13.4^{+6.4}_{-2.9}$ | $13.1^{+6.6}_{-2.7}$ | $13.0^{+6.6}_{-2.6}$ | $13.1^{+6.6}_{-2.7}$ | ... | ... |
| | Age (Gyr) (A09) | $2.6^{+4.0}_{-1.0}$ | $2.5^{+3.8}_{-1.0}$ | $2.6^{+4.5}_{-1.0}$ | $2.8^{+3.8}_{-1.2}$ | $2.6^{+3.8}_{-1.0}$ | $2.6^{+3.9}_{-1.0}$ | $2.6^{+3.9}_{-1.0}$ | ... | ... |
| | Age (Gyr) (J10) | $2.0^{+4.4}_{-1.0}$ | $1.9^{+4.1}_{-0.9}$ | $2.0^{+5.0}_{-1.0}$ | $2.1^{+4.2}_{-1.1}$ | $2.0^{+4.2}_{-1.0}$ | $1.9^{+4.2}_{-0.9}$ | $2.0^{+4.3}_{-1.0}$ | ... | ... |
| | Age (Gyr) (S21) | $3.3^{+5.3}_{-1.3}$ | $3.2^{+5.0}_{-1.3}$ | $3.4^{+6.0}_{-1.4}$ | $3.5^{+5.1}_{-1.5}$ | $3.4^{+5.1}_{-1.4}$ | $3.2^{+5.2}_{-1.3}$ | $3.3^{+5.3}_{-1.4}$ | ... | ... |
| | Median age (Gyr) (simulation) | $4.1^{+2.8}_{-2.7}$ | $4.1^{+2.8}_{-2.7}$ | $4.1^{+2.7}_{-2.7}$ | $4.3^{+2.6}_{-2.8}$ | $4.1^{+2.8}_{-2.7}$ | $4.2^{+2.6}_{-2.8}$ | $4.2^{+2.7}_{-2.7}$ | ... | ... |
| | K-S (A09 simulation) | 0.2 | 0.3 | 0.2 | 0.2 | 0.2 | 0.3 | ... | ... | ... |
| | K-S (J10 simulation) | 0.3 | 0.3 | 0.3 | 0.3 | 0.3 | 0.3 | ... | ... | ... |
| | K-S (S21 simulation) | 0.2 | 0.2 | 0.2 | 0.2 | 0.2 | 0.2 | ... | ... | ... |
| L | $H$ (pc) | $173^{+50}_{-35}$ | $157^{+39}_{-29}$ | $134^{+26}_{-20}$ | $136^{+27}_{-21}$ | $143^{+31}_{-23}$ | $210^{+98}_{-52}$ | $153^{+56}_{-30}$ | 18 | 7 |
| | $\sigma_w$ (km s$^{-1}$) | $12.6^{+1.7}_{-1.3}$ | $12.0^{+1.4}_{-1.2}$ | $11.1^{+1.0}_{-0.9}$ | $11.2^{+1.1}_{-0.9}$ | $11.5^{+1.2}_{-1.0}$ | $13.9^{+2.9}_{-1.8}$ | $11.8^{+2.0}_{-1.2}$ | ... | ... |
| | Age (Gyr) (A09) | $2.4^{+0.8}_{-0.5}$ | $2.2^{+0.6}_{-0.4}$ | $1.8^{+0.4}_{-0.3}$ | $1.8^{+0.4}_{-0.3}$ | $1.9^{+0.5}_{-0.4}$ | $3.0^{+1.6}_{-0.8}$ | $2.1^{+0.9}_{-0.5}$ | ... | ... |
| | Age (Gyr) (J10) | $1.8^{+0.8}_{-0.5}$ | $1.6^{+0.6}_{-0.4}$ | $1.2^{+0.4}_{-0.3}$ | $1.3^{+0.4}_{-0.3}$ | $1.4^{+0.5}_{-0.3}$ | $2.4^{+1.7}_{-0.8}$ | $1.5^{+0.9}_{-0.4}$ | ... | ... |
| | Age (Gyr) (S21) | $3.0^{+1.0}_{-0.7}$ | $2.7^{+0.8}_{-0.6}$ | $2.3^{+0.5}_{-0.4}$ | $2.3^{+0.6}_{-0.4}$ | $2.4^{+0.6}_{-0.5}$ | $3.8^{+2.1}_{-1.1}$ | $2.6^{+1.2}_{-0.6}$ | ... | ... |
| | Median age (Gyr) (simulation) | $2.1^{+3.7}_{-1.6}$ | $2.3^{+3.8}_{-1.7}$ | $2.3^{+3.7}_{-1.8}$ | $2.7^{+3.6}_{-2.1}$ | $2.3^{+3.8}_{-1.8}$ | $1.6^{+3.4}_{-1.2}$ | $2.2^{+3.8}_{-1.7}$ | ... | ... |
| | K-S (A09 simulation) | 0.4 | 0.3 | 0.4 | 0.5 | 0.4 | 0.5 | ... | ... | ... |
| | K-S (J10 simulation) | 0.3 | 0.4 | 0.5 | 0.6 | 0.5 | 0.4 | ... | ... | ... |
| | K-S (S21 simulation) | 0.4 | 0.4 | 0.4 | 0.4 | 0.4 | 0.6 | ... | ... | ... |
| T | $H$ (pc) | $170^{+143}_{-52}$ | $173^{+150}_{-55}$ | $179^{+158}_{-59}$ | $173^{+139}_{-55}$ | $183^{+156}_{-62}$ | $170^{+146}_{-53}$ | $175^{+149}_{-56}$ | 4 | 0.4 |
| | $\sigma_w$ (km s$^{-1}$) | $12.5^{+4.5}_{-2.1}$ | $12.6^{+4.6}_{-2.2}$ | $12.8^{+4.8}_{-2.3}$ | $12.6^{+4.3}_{-2.2}$ | $13.0^{+4.7}_{-2.4}$ | $12.5^{+4.5}_{-2.1}$ | $12.7^{+4.6}_{-2.2}$ | ... | ... |

**Table 7**
*(Continued)*

| SpT | Quantity | B97 | B01 | B03 | SM08 | M18 | P20 | Median[a] | $N_{obs}$[b] | $N_{thick}$[c] |
|---|---|---|---|---|---|---|---|---|---|---|
| | Age (Gyr) (A09) | $2.3^{+2.3}_{-0.8}$ | $2.4^{+2.4}_{-0.8}$ | $2.5^{+2.6}_{-0.9}$ | $2.4^{+2.2}_{-0.8}$ | $2.5^{+2.5}_{-1.0}$ | $2.4^{+2.4}_{-0.8}$ | $2.4^{+2.4}_{-0.8}$ | ⋯ | ⋯ |
| | Age (Gyr) (J10) | $1.7^{+2.4}_{-0.7}$ | $1.8^{+2.6}_{-0.8}$ | $1.9^{+2.7}_{-0.8}$ | $1.8^{+2.4}_{-0.8}$ | $1.9^{+2.7}_{-0.9}$ | $1.8^{+2.5}_{-0.8}$ | $1.8^{+2.5}_{-0.8}$ | ⋯ | ⋯ |
| | Age (Gyr) (S21) | $3.0^{+3.1}_{-1.0}$ | $3.0^{+3.2}_{-1.1}$ | $3.2^{+3.3}_{-1.2}$ | $3.0^{+3.4}_{-1.1}$ | $3.2^{+3.4}_{-1.2}$ | $3.0^{+3.1}_{-1.1}$ | $3.1^{+3.2}_{-1.1}$ | ⋯ | ⋯ |
| | Median age (Gyr) (simulation) | $3.6^{+3.0}_{-2.4}$ | $3.5^{+3.0}_{-2.4}$ | $3.7^{+3.0}_{-2.5}$ | $3.6^{+3.0}_{-2.4}$ | $3.6^{+3.0}_{-2.5}$ | $3.7^{+3.0}_{-2.6}$ | $3.6^{+3.0}_{-2.5}$ | ⋯ | ⋯ |
| | K-S (A09 simulation) | 0.3 | 0.3 | 0.2 | 0.3 | 0.2 | 0.3 | ⋯ | ⋯ | ⋯ |
| | K-S (J10 simulation) | 0.4 | 0.3 | 0.3 | 0.3 | 0.3 | 0.4 | ⋯ | ⋯ | ⋯ |
| | K-S (S21 simulation) | 0.2 | 0.2 | 0.2 | 0.2 | 0.2 | 0.2 | ⋯ | ⋯ | ⋯ |

**Notes.** Model notation—B97: Burrows et al. (1997); B01: Burrows et al. (2001); B03: Baraffe et al. (2003); SM08: Saumon & Marley (2008); M18: Marley et al. (2018); P20: Phillips et al. (2020).
[a] Median values computed by combining all samples obtained from all models, not by standard error propagation formula. For M7–M9 dwarfs, only the B97, B01, and B03 models are used; for L0–L4 dwarfs, the P20 models are excluded.
[b] The total number of UCDs in the magnitude-limited WISP and 3D-HST surveys analyzed in this study; simulations are compared to $N_{thin} = N_{obs} - N_{thick}$.
[c] Estimated number of thick disk contaminants; see Section 3.4.
[d] Estimated probability that the age distribution of the simulated sample and that inferred from our measured number counts are distinct, as assessed through the K-S test. Values of K-S ⩾ 0.7 are considered significantly distinct.

of these objects. Our results are in qualitative agreement with the predictions of Burgasser (2004) and Ryan et al. (2017), who argued that brown dwarf evolutionary effects would drive down the vertical scaleheight of late-L and early T dwarfs compared to late-M dwarfs. The large uncertainties on our scaleheights for these cooler objects, driven by the small sample of L and T dwarfs identified, limit our ability to explore these differences with this sample.

### 4.3. Velocity Dispersions

The scaleheight, velocity dispersion, and age of a stellar or substellar population in the Galaxy are interdependent. Older stars tend to have larger kinematic dispersions and occupy a disk with a larger scaleheight (Sanders & Das 2018; Bovy 2017). This trend can be explained by cumulative dynamic interactions with structures in the Galaxy such as giant molecular clouds or spiral arms (Spitzer & Schwarzschild 1953; Lacey 1984; Sellwood & Binney 2002), or dynamical heating due to past merger events (Toth & Ostriker 1992; Hopkins et al. 2008; Martig et al. 2014; Minchev et al. 2015; Ma et al. 2017; Buck et al. 2020).

To infer population ages for each of our spectral subgroups, we first converted scaleheights into total velocity dispersions using a relationship based on the analytical model of van der Kruit (1988),

$$H = \zeta_n \frac{\sigma_{20}^2}{\Sigma_{68}}, \quad (17)$$

where $\sigma_{20}$ is the vertical velocity dispersion in units of 20 km s$^{-1}$, $\Sigma_{68}$ is the surface mass density of the Galactic disk in units of 68 $M_\odot$ pc$^{-2}$ (Bovy & Rix 2013), and $\zeta_n$ is a normalization constant based on the parametric form of the vertical mass density $\rho(z)$:

$$\rho_n(z) = 2^{-2/n} \rho_0 \, \text{sech}^{2/n}\left(\frac{nz}{2H}\right). \quad (18)$$

Our simple exponential disk corresponds to $n = \infty$ and $\zeta_n = 435$ pc (Ryan et al. 2017).

Converting our scaleheights into velocity dispersions using these relations (Table 7), we compared these inferred dispersions to dispersion measurements from local UCD samples ($d < 20$ pc). Our median estimate of $\sigma_W = 15.1^{+1.4}_{-2.0}$ km s$^{-1}$ for M7–M9 dwarfs is consistent with dispersions reported by Burgasser et al. (2015, $\sigma_W = 13.8 \pm 0.3$ km s$^{-1}$) and Hsu et al. (2021, $\sigma_W = 16.3 \pm 0.3$ km s$^{-1}$). In contrast, our median estimate of $\sigma_W = 11.6^{+1.5}_{-1.1}$ km s$^{-1}$ for L0–L4 dwarfs is significantly lower than the measured dispersion reported by Burgasser et al. (2015, $\sigma_W = 19.5 \pm 0.4$ km s$^{-1}$) for this subtype group. In this case, there is evidence that the local population of L dwarfs is significantly contaminated by older, thick disk sources, and Hsu et al. (2021) report a smaller thin disk L dwarf dispersion, $\sigma_W = 15.3 \pm 0.3$ km s$^{-1}$, which is still higher than our estimate of $\sigma_W = 11.8^{+2.0}_{-1.2}$ km s$^{-1}$ for our overall L dwarf sample. We find agreement between our median estimate of $\sigma_W = 12.7^{+4.6}_{-2.2}$ km s$^{-1}$ for T dwarfs and the local measurement of $\sigma_W = 13.3 \pm 0.4$ km s$^{-1}$ by Hsu et al. (2021), albeit with substantial uncertainties in our value due to small number statistics.

### 4.4. Ages

We transformed velocity dispersions into ages using a power-law age-velocity dispersion relation (AVR) of the form;

$$\sigma_W(\tau) = \sigma_0 \left(\frac{\tau + \tau_{min}}{\tau_{min} + \tau_{max}}\right)^{\beta_z}, \quad (19)$$

where $\sigma_0$ is the velocity dispersion at $\tau_{max}$, $\tau_{min}$, and $\tau_{max}$ define the epochs of initial and maximal velocity dispersions, and $\beta_z$ is a power-law index that quantifies the rate of dispersion increase. We used the best-fit model parameters from Aumer & Binney (2009) without a low-metallicity tail: $(\beta, \sigma_0, \tau_{min}, \tau_{max}) = (0.445, 23.831 \text{ km s}^{-1}, 0.001 \text{ Gyr}, 10 \text{ Gyr})$.

Inferred ages and associated uncertainty distributions (based on Monte Carlo sampling) for each spectral subgroup and model are listed in Table 7 and illustrated in Figure 12. In the figure, we also show the distribution of the median age from the original simulations. To estimate this distribution, we randomly sampled $N_{thin}$ objects from the simulation for a given spectral type range and computed the median age, where $N_{thin}$ corresponds to the observed sample $N_{obs}$ minus the estimated thick disk contamination $N_{thick}$ (Tables 7 and 8). This process was repeated $10^4$ times to estimate the sampling uncertainty in the simulated median age. Different evolutionary models display similar patterns in age versus spectral type grouping, with M7–M9 dwarfs (mostly stars) and T5–Y1 dwarfs (all brown dwarfs) having near-uniform ages, while L0–L4 dwarfs (mixed stars and brown dwarfs) are skewed toward younger ages.

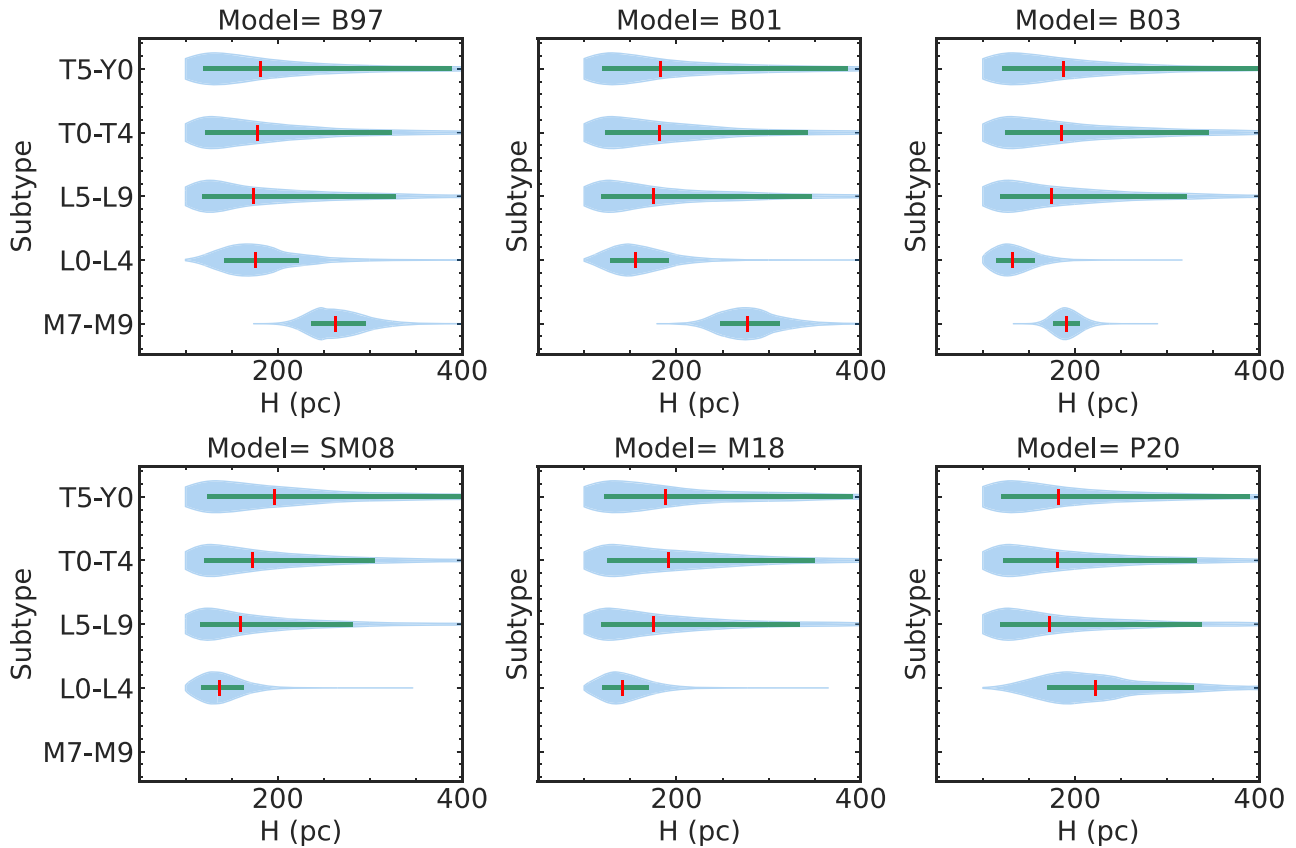

**Figure 11.** Probability distributions of scaleheights inferred from our sample (blue violin plots) based on Poisson errors and scaleheight/number count relation. Median values are indicated by vertical red bars and their 16%–84% quantile range is displayed as green horizontal bars. The extended shapes of these distributions for spectral types >L5 reflect the intrinsic sampling uncertainty from the small numbers of these bins.

To test more generally whether the distributions of simulated ages agree with age distributions inferred from number counts, we computed the overlapping probability using the K-S test, which measures the probability that two distributions are distinct. We find K-S values are generally ⩽0.5, indicating good to modest agreement between simulated and inferred age distributions. There are significant deviations between simulated and inferred ages for M7–M9 and L0–L4 subtypes, which may indicate errors in the evolutionary models themselves, with some models having smaller inferred ages compared to simulated ages (B03, SM08, M18) and others larger ages (B97, B01, P20). These systematic differences between the models are worse when a lower thick disk fraction is considered (Table 8), favoring a 12% fraction. We note that the largest deviation between inferred and simulated ages occurs for the P20 models applied to the L0–L4 dwarf sample, and for all models applied to the M5–M9 sample. We attribute these mismatches to the parameter limits in the models, an incorrect assumption of the thick disk fraction or systematic uncertainties in age-velocity dispersion relations. Nevertheless, we find reasonable agreement between the inferred and the median simulated ages for L and T dwarfs, albeit with significant uncertainties due to small number statistics among the latest spectral class groups.

Our age estimates for UCDs based on scaleheights can be compared to kinematic age estimates from the local population. Our median M7–M9 dwarf age estimate of $3.6^{+0.8}_{-1.0}$ Gyr based on the Aumer & Binney (2009) AVR is consistent with kinematic estimates from Burgasser et al. (2015, $4.0 \pm 0.2$ Gyr) and Hsu et al. (2021, $4.1 \pm 0.3$ Gyr).

For L0–L4 dwarfs, our estimate of $2.0^{+0.6}_{-0.4}$ Gyr is substantially lower than the (likely contaminated) $6.5 \pm 0.4$ Gyr kinematic age inferred by Burgasser et al. (2015), and our age estimate for all L dwarfs, $2.1^{+0.9}_{-0.5}$ Gyr, is also significantly lower than the kinematic age of thin disk L dwarfs of $4.2 \pm 0.3$ Gyr from Hsu et al. (2021). Our estimate of $2.4^{+2.4}_{-0.8}$ Gyr for T dwarfs is lower than but formally in agreement with the kinematic age of $3.5 \pm 0.3$ Gyr from Hsu et al. (2021), albeit again with significant uncertainties.

The absolute kinematic ages reported here and in other population studies must be considered carefully, as there are significant systematic issues to be considered when mapping spatial or kinematic distributions to population ages. Generally, these arise from the fact that populations formed in distinct environments and residing in different regions of the Galaxy undergo different kinematic heating histories (Bovy et al. 2012; Aumer et al. 2016; Mackereth et al. 2019; Sharma et al. 2021). Relations connecting quantities such as age, velocity dispersion, and spatial distribution measured locally may not apply to populations located at large vertical distances from the Galactic plane or in different radial zones. More succinctly, the AVRs derived from different stellar samples do have significant variations. The AVR of Aumer & Binney (2009), which is used in prior UCD population analyses (Burgasser et al. 2015; Ryan et al. 2017; Hsu et al. 2021), is based on ≈15,000 main-sequence stars with distances up to ∼300 pc from the Sun selected from the Geneva Copenhagen Survey (GCS; Nordström et al. 2004; Holmberg et al. 2007) and Hipparcos (van Leeuwen 2007) catalogs, with isochronal ages based on the (Bertelli et al. 2008) models. We examined the AVRs of two other stellar samples to examine systematic effects, both of which use a model of the same form as Equation 19. The AVR of Just & Jahreiß (2010, hereafter J10) is based on a sample of main-sequence GCS and Hipparcos stars, with additional sources from Jahreiß & Wielen (1997), sampling distances up to ∼200 pc from the Sun. Following a similar method as A09, J10 model the observed number counts, kinematics, ages, and metallicities assuming an initial mass function, star formation rate, and metallicity distribution in a fully consistent model of the Galaxy, but with different parameterizations of the star formation rate and gravitational potential. The resulting AVR best-fit parameters are $(\beta, \sigma_0, \tau_{min}, \tau_{max}) = (0.5, 29.9 \text{ km s}^{-1}, 0.5 \text{ Gyr}, 12 \text{ Gyr})$. The AVR of Sharma et al. (2021, hereafter S21) is based on a large sample of ∼840,000 main-sequence turnoff and red-giant stars out to distances of 2 kpc from the Sun (Sharma et al. 2018, 2019), with measurements from GALAH (De Silva et al. 2015), LAMOST (Zhao et al. 2012), APOGEE (Hayden et al. 2015), the Transit Exoplanet Survey Satellite (Ricker et al. 2014) catalog, the High Efficiency and Resolution Multi-Element Spectrograph (Sheinis et al. 2015), and the Kepler/K2 mission (Borucki et al. 2010; Howell et al. 2014). Ages were derived from isochronal fitting and astroseismology, and the resulting AVR best-fit parameters are $(\beta, \sigma_0, \tau_{min}, \tau_{max}) = (0.441 \pm 0.007, 21.1 \pm 0.2 \text{ km s}^{-1}, 0.1 \text{ Gyr}, 10 \text{ Gyr})$. As indicated in Tables 7 and 8, and illustrated in Figure 13, the J10 relation consistently yields ages that are older than A09 by ∼1–1.5 Gyr while the S21 relation consistently yields ages that are younger by ∼0.5–1 Gyr, implying an overall systematic uncertainty on absolute ages of 1–2 Gyr. Despite this, all three relations predict the same relative trends in inferred ages across UCD spectral subgroups; they all find that the late-M dwarfs are older than the L and T dwarfs. Thus, while the absolute ages of UCD populations may be uncertain, the relative age differences predicted by the simulations are confirmed by our scaleheight measurements.

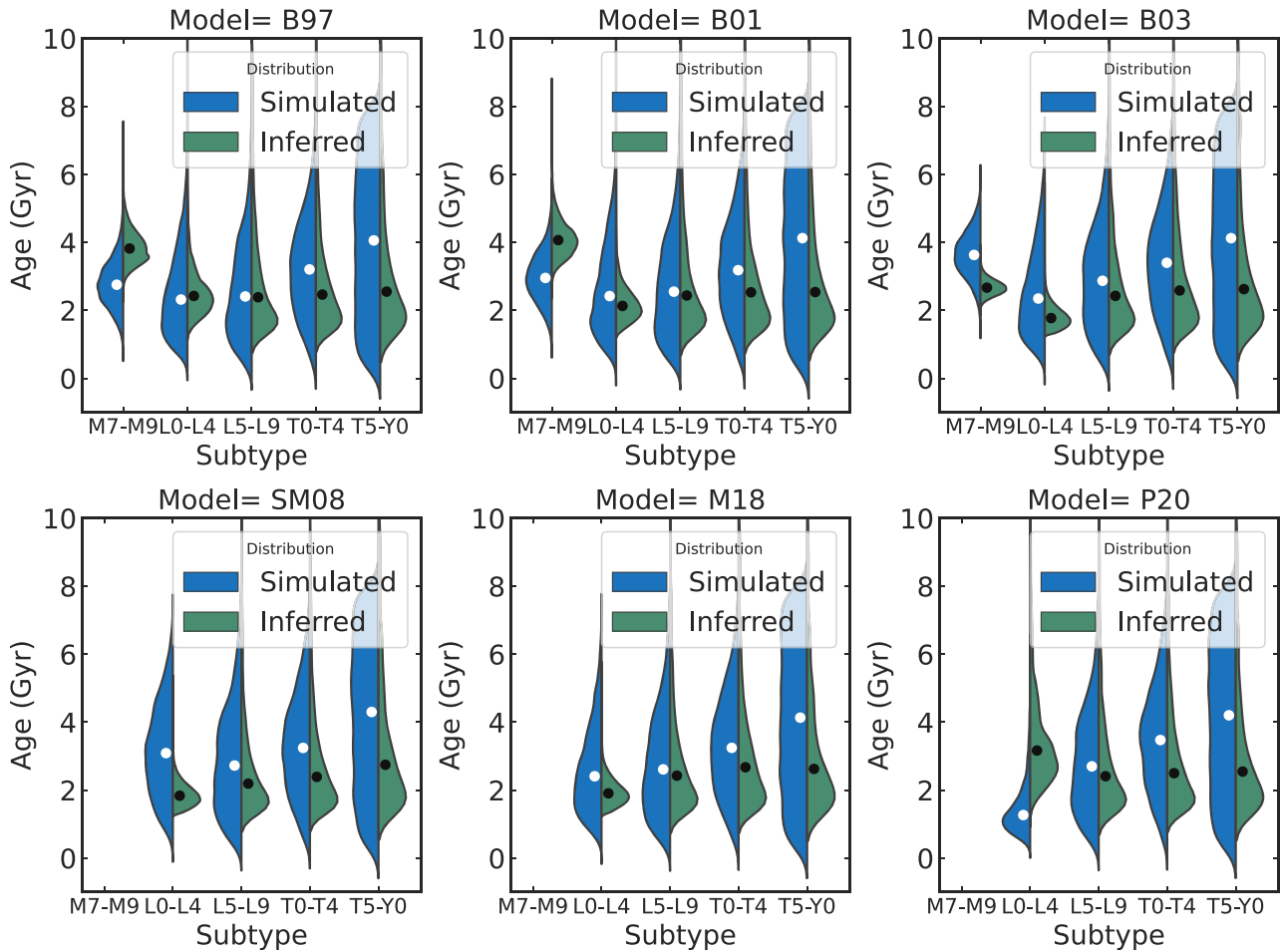

**Figure 12.** Comparison between the distribution of simulated median ages (blue shaded regions on left) and the median age estimates for our HST UCD sample based on scaleheight determinations (olive shaded regions on right) and the Aumer & Binney (2009) AVR. Median values for each distribution are indicated by white and black circles, respectively. Note that the SM08, M18, and P20 number count predictions are unreliable for M7–M9 dwarfs due to model parameter limits, and are not displayed.

## 5. Predictions for the JWST PASSAGE Survey

In this series, we have examined the largest deep UCD spectral sample compiled to date, composed of 164 late-M, L, and T dwarfs with 1.1–1.7 μm low-resolution spectra. While reaching kiloparsec distances for the warmest late-M and L

Table 8
Scaleheights, Velocity Dispersions, and Population Ages of HST UCDs Assuming a Thick Disk Fraction of 5%

| SpT | Quantity | B97 | B01 | B03 | SM08 | M18 | P20 | Median[a] | $N_{\rm obs}$[b] | $N_{\rm thick}$[c] |
|---|---|---|---|---|---|---|---|---|---|---|
| M7–M9 | $H$ (pc) | $385^{+72}_{-53}$ | $429^{+108}_{-73}$ | $240^{+22}_{-18}$ | ⋯ | ⋯ | ⋯ | $361^{+112}_{-125}$ | 76 | 15.3 |
| | $\sigma_w$ (km s$^{-1}$) | $18.8^{+1.7}_{-1.3}$ | $19.9^{+2.4}_{-1.8}$ | $14.8^{+0.7}_{-0.6}$ | ⋯ | ⋯ | ⋯ | $18.2^{+2.6}_{-3.5}$ | ⋯ | ⋯ |
| | Age (Gyr) (A09) | $5.9^{+1.2}_{-0.9}$ | $6.6^{+1.9}_{-1.3}$ | $3.4^{+0.4}_{-0.3}$ | ⋯ | ⋯ | ⋯ | $5.5^{+1.9}_{-2.1}$ | ⋯ | ⋯ |
| | Age (Gyr) (J10) | $5.5^{+1.5}_{-1.0}$ | $6.4^{+2.3}_{-1.4}$ | $2.9^{+0.4}_{-0.3}$ | ⋯ | ⋯ | ⋯ | $5.1^{+2.3}_{-2.3}$ | ⋯ | ⋯ |
| | Age (Gyr) (S21) | $7.7^{+1.7}_{-1.2}$ | $8.7^{+2.6}_{-1.7}$ | $4.4^{+0.5}_{-0.4}$ | ⋯ | ⋯ | ⋯ | $7.1^{+2.6}_{-2.8}$ | ⋯ | ⋯ |
| | Median age (Gyr) (simulation) | $2.8^{+0.6}_{-0.6}$ | $3.0^{+0.7}_{-0.6}$ | $3.6^{+0.6}_{-0.6}$ | ⋯ | ⋯ | ⋯ | $2.7^{+1.0}_{-2.4}$ | | |
| | K-S (A09 simulation) | **1.0** | **1.0** | 0.2 | ⋯ | ⋯ | ⋯ | ⋯ | ⋯ | |
| | K-S (J10 simulation) | **1.0** | **1.0** | 0.6 | ⋯ | ⋯ | ⋯ | ⋯ | ⋯ | |
| | K-S (S21 simulation) | **1.0** | **1.0** | 0.6 | ⋯ | ⋯ | ⋯ | ⋯ | ⋯ | |
| L0–L4 | $H$ (pc) | $235^{+104}_{-53}$ | $199^{+59}_{-38}$ | $161^{+31}_{-26}$ | $168^{+35}_{-28}$ | $177^{+42}_{-31}$ | $344^{+238}_{-113}$ | $182^{+63}_{-36}$ | 16 | 2.9 |
| | $\sigma_w$ (km s$^{-1}$) | $14.7^{+3.0}_{-1.8}$ | $13.5^{+1.9}_{-1.3}$ | $12.2^{+1.1}_{-1.1}$ | $12.4^{+1.2}_{-1.0}$ | $12.8^{+1.4}_{-1.2}$ | $17.8^{+5.3}_{-3.2}$ | $13.0^{+2.1}_{-1.4}$ | ⋯ | ⋯ |
| | Age (Gyr) (A09) | $3.4^{+1.7}_{-0.8}$ | $2.8^{+1.0}_{-0.6}$ | $2.2^{+0.5}_{-0.4}$ | $2.3^{+0.5}_{-0.4}$ | $2.4^{+0.7}_{-0.5}$ | $5.2^{+4.2}_{-1.9}$ | $2.5^{+1.0}_{-0.6}$ | ⋯ | ⋯ |
| | Age (Gyr) (J10) | $2.8^{+1.8}_{-0.8}$ | $2.2^{+1.0}_{-0.6}$ | $1.6^{+0.5}_{-0.4}$ | $1.7^{+0.5}_{-0.4}$ | $1.8^{+0.7}_{-0.5}$ | $4.7^{+5.0}_{-2.0}$ | $1.9^{+1.0}_{-0.5}$ | ⋯ | ⋯ |
| | Age (Gyr) (S21) | $4.4^{+2.3}_{-1.1}$ | $3.6^{+1.3}_{-0.8}$ | $2.8^{+0.7}_{-0.5}$ | $2.9^{+0.7}_{-0.6}$ | $3.1^{+0.9}_{-0.6}$ | $6.8^{+5.6}_{-2.5}$ | $3.2^{+1.4}_{-0.8}$ | ⋯ | ⋯ |
| | Median age (Gyr) (simulation) | $2.3^{+1.4}_{-1.0}$ | $2.5^{+1.3}_{-1.0}$ | $2.4^{+0.7}_{-1.0}$ | $3.1^{+1.3}_{-1.3}$ | $2.4^{+1.4}_{-1.0}$ | $1.3^{+0.7}_{-0.4}$ | $2.5^{+1.4}_{-1.1}$ | ⋯ | |
| | K-S (A09 simulation) | 0.4 | 0.3 | 0.3 | 0.4 | 0.2 | **0.9** | ⋯ | ⋯ | ⋯ |
| | K-S (J10 simulation) | 0.2 | 0.2 | 0.5 | 0.6 | 0.3 | **0.9** | ⋯ | ⋯ | ⋯ |
| | K-S (S21 simulation) | 0.6 | 0.5 | 0.3 | 0.2 | 0.4 | **1.0** | ⋯ | ⋯ | ⋯ |
| L5–L9 | $H$ (pc) | $180^{+189}_{-62}$ | $183^{+210}_{-65}$ | $182^{+172}_{-63}$ | $166^{+155}_{-51}$ | $181^{+191}_{-62}$ | $180^{+202}_{-62}$ | $179^{+189}_{-62}$ | 2 | 0.1 |
| | $\sigma_w$ (km s$^{-1}$) | $12.9^{+5.6}_{-2.5}$ | $13.0^{+6.0}_{-2.5}$ | $12.9^{+5.1}_{-2.5}$ | $12.4^{+4.8}_{-2.5}$ | $12.9^{+5.6}_{-2.5}$ | $12.8^{+5.9}_{-2.5}$ | $12.8^{+5.6}_{-2.4}$ | ⋯ | ⋯ |
| | Age (Gyr) (A09) | $2.5^{+3.1}_{-1.0}$ | $2.6^{+3.5}_{-1.0}$ | $2.5^{+2.8}_{-1.0}$ | $2.3^{+2.5}_{-0.8}$ | $2.5^{+3.1}_{-1.0}$ | $2.5^{+3.3}_{-1.0}$ | $2.5^{+3.1}_{-0.9}$ | ⋯ | ⋯ |
| | Age (Gyr) (J10) | $1.9^{+3.3}_{-0.9}$ | $2.0^{+3.7}_{-0.9}$ | $1.9^{+3.0}_{-0.9}$ | $1.7^{+2.6}_{-0.7}$ | $1.9^{+3.4}_{-0.9}$ | $1.9^{+3.6}_{-0.9}$ | $1.9^{+3.3}_{-0.9}$ | ⋯ | ⋯ |
| | Age (Gyr) (S21) | $3.2^{+4.2}_{-1.3}$ | $3.3^{+4.6}_{-1.3}$ | $3.2^{+3.8}_{-1.3}$ | $2.9^{+3.3}_{-1.0}$ | $3.2^{+4.2}_{-1.3}$ | $3.2^{+4.4}_{-1.3}$ | $3.2^{+4.1}_{-1.2}$ | ⋯ | ⋯ |
| | Median age (Gyr) (simulation) | $2.4^{+1.8}_{-1.3}$ | $2.6^{+1.8}_{-1.4}$ | $2.8^{+1.8}_{-1.6}$ | $2.7^{+1.9}_{-1.5}$ | $2.6^{+1.8}_{-1.4}$ | $2.6^{+1.8}_{-1.4}$ | $2.6^{+1.8}_{-1.4}$ | ⋯ | |
| | K-S (A09 simulation) | 0.2 | 0.2 | 0.1 | 0.2 | 0.2 | 0.2 | ⋯ | ⋯ | ⋯ |
| | K-S (J10 simulation) | 0.1 | 0.1 | 0.2 | 0.3 | 0.2 | 0.2 | ⋯ | ⋯ | ⋯ |
| | K-S (S21 simulation) | 0.3 | 0.3 | 0.2 | 0.2 | 0.2 | 0.2 | ⋯ | ⋯ | ⋯ |
| T0–T4 | $H$ (pc) | $191^{+189}_{-68}$ | $197^{+215}_{-72}$ | $202^{+204}_{-76}$ | $188^{+169}_{-66}$ | $209^{+212}_{-81}$ | $195^{+189}_{-71}$ | $197^{+197}_{-72}$ | 3 | 0.3 |
| | $\sigma_w$ (km s$^{-1}$) | $13.3^{+5.4}_{-2.6}$ | $13.5^{+6.0}_{-2.8}$ | $13.6^{+5.7}_{-2.8}$ | $13.1^{+5.0}_{-2.6}$ | $13.9^{+5.8}_{-3.0}$ | $13.4^{+5.4}_{-2.7}$ | $13.4^{+5.6}_{-2.8}$ | ⋯ | ⋯ |
| | Age (Gyr) (A09) | $2.7^{+3.1}_{-1.1}$ | $2.8^{+3.6}_{-1.1}$ | $2.8^{+3.4}_{-1.2}$ | $2.6^{+2.8}_{-1.0}$ | $3.0^{+3.6}_{-1.3}$ | $2.7^{+3.1}_{-1.1}$ | $2.8^{+3.3}_{-1.1}$ | ⋯ | ⋯ |
| | Age (Gyr) (J10) | $2.1^{+3.4}_{-1.1}$ | $2.2^{+3.9}_{-1.1}$ | $2.2^{+3.7}_{-1.1}$ | $2.0^{+3.0}_{-1.0}$ | $2.4^{+3.9}_{-1.2}$ | $2.1^{+3.4}_{-1.0}$ | $2.2^{+3.6}_{-1.1}$ | ⋯ | ⋯ |
| | Age (Gyr) (S21) | $3.4^{+4.1}_{-1.5}$ | $3.6^{+4.8}_{-1.5}$ | $3.7^{+4.5}_{-1.6}$ | $3.4^{+3.7}_{-1.3}$ | $3.8^{+4.7}_{-1.7}$ | $3.5^{+4.2}_{-1.4}$ | $3.5^{+4.4}_{-1.4}$ | ⋯ | ⋯ |
| | Median age (Gyr) (simulation) | $2.9^{+2.0}_{-1.4}$ | $2.9^{+2.1}_{-1.5}$ | $3.2^{+2.1}_{-1.7}$ | $3.0^{+2.0}_{-1.6}$ | $3.0^{+2.0}_{-1.6}$ | $3.2^{+2.1}_{-1.6}$ | $3.0^{+2.1}_{-1.6}$ | ⋯ | |
| | K-S (A09 simulation) | 0.1 | 0.1 | 0.1 | 0.1 | 0.1 | 0.1 | ⋯ | ⋯ | ⋯ |
| | K-S (J10 simulation) | 0.2 | 0.2 | 0.2 | 0.2 | 0.2 | 0.3 | ⋯ | ⋯ | ⋯ |
| | K-S (S21 simulation) | 0.2 | 0.2 | 0.2 | 0.2 | 0.2 | 0.2 | ⋯ | ⋯ | ⋯ |
| T5–Y0 | $H$ (pc) | $189^{+247}_{-70}$ | $189^{+246}_{-71}$ | $193^{+277}_{-73}$ | $204^{+235}_{-81}$ | $194^{+238}_{-73}$ | $187^{+238}_{-69}$ | $193^{+247}_{-73}$ | 1 | 0.1 |
| | $\sigma_w$ (km s$^{-1}$) | $13.2^{+6.8}_{-2.7}$ | $13.2^{+6.8}_{-2.7}$ | $13.3^{+7.5}_{-2.8}$ | $13.7^{+6.4}_{-3.1}$ | $13.4^{+6.6}_{-2.8}$ | $13.1^{+6.6}_{-2.7}$ | $13.3^{+6.8}_{-2.8}$ | ⋯ | ⋯ |
| | Age (Gyr) (A09) | $2.6^{+4.1}_{-1.1}$ | $2.6^{+4.1}_{-1.1}$ | $2.7^{+4.7}_{-1.1}$ | $2.9^{+3.9}_{-1.2}$ | $2.7^{+4.0}_{-1.1}$ | $2.6^{+4.0}_{-1.0}$ | $2.7^{+4.1}_{-1.1}$ | ⋯ | ⋯ |
| | Age (Gyr) (J10) | $2.0^{+4.5}_{-1.0}$ | $2.0^{+4.5}_{-1.0}$ | $2.1^{+5.2}_{-1.1}$ | $2.3^{+4.4}_{-1.2}$ | $2.1^{+4.4}_{-1.1}$ | $2.0^{+4.3}_{-1.0}$ | $2.1^{+4.5}_{-1.1}$ | ⋯ | ⋯ |
| | Age (Gyr) (S21) | $3.4^{+5.5}_{-1.4}$ | $3.4^{+5.5}_{-1.4}$ | $3.5^{+6.2}_{-1.5}$ | $3.7^{+5.2}_{-1.7}$ | $3.5^{+5.3}_{-1.5}$ | $3.3^{+5.3}_{-1.4}$ | $3.5^{+5.5}_{-1.5}$ | ⋯ | ⋯ |
| | Median age (Gyr) (simulation) | $4.2^{+2.7}_{-2.8}$ | $4.1^{+2.8}_{-2.6}$ | $4.1^{+2.8}_{-2.6}$ | $4.3^{+2.7}_{-2.8}$ | $4.2^{+2.7}_{-2.7}$ | $4.1^{+2.7}_{-2.7}$ | $4.1^{+2.7}_{-2.7}$ | ⋯ | |
| | K-S (A09 simulation) | 0.2 | 0.2 | 0.2 | 0.2 | 0.2 | 0.2 | ⋯ | ⋯ | ⋯ |
| | K-S (J10 simulation) | 0.3 | 0.3 | 0.3 | 0.3 | 0.3 | 0.3 | ⋯ | ⋯ | ⋯ |
| | K-S (S21 simulation) | 0.2 | 0.2 | 0.2 | 0.2 | 0.2 | 0.2 | ⋯ | ⋯ | ⋯ |
| L | $H$ (pc) | $227^{+87}_{-50}$ | $198^{+56}_{-37}$ | $162^{+31}_{-26}$ | $166^{+33}_{-27}$ | $176^{+41}_{-31}$ | $313^{+199}_{-96}$ | $190^{+100}_{-41}$ | 18 | 3 |
| | $\sigma_w$ (km s$^{-1}$) | $14.5^{+2.5}_{-1.7}$ | $13.5^{+1.8}_{-1.3}$ | $12.2^{+1.1}_{-1.0}$ | $12.3^{+1.2}_{-1.0}$ | $12.7^{+1.4}_{-1.2}$ | $17.0^{+4.7}_{-2.8}$ | $13.2^{+3.1}_{-1.5}$ | ⋯ | ⋯ |
| | Age (Gyr) (A09) | $3.2^{+1.4}_{-0.8}$ | $2.8^{+0.9}_{-0.6}$ | $2.2^{+0.5}_{-0.4}$ | $2.3^{+0.5}_{-0.4}$ | $2.4^{+0.6}_{-0.5}$ | $4.7^{+3.4}_{-1.6}$ | $2.7^{+1.6}_{-0.6}$ | ⋯ | ⋯ |
| | Age (Gyr) (J10) | $2.7^{+1.5}_{-0.8}$ | $2.2^{+0.9}_{-0.6}$ | $1.6^{+0.5}_{-0.4}$ | $1.7^{+0.5}_{-0.4}$ | $1.8^{+0.6}_{-0.4}$ | $4.2^{+4.0}_{-1.7}$ | $2.0^{+1.7}_{-0.6}$ | ⋯ | ⋯ |
| | Age (Gyr) (S21) | $4.2^{+1.9}_{-1.0}$ | $3.6^{+1.2}_{-0.8}$ | $2.8^{+0.6}_{-0.5}$ | $2.9^{+0.7}_{-0.6}$ | $3.1^{+0.9}_{-0.6}$ | $6.1^{+4.6}_{-2.1}$ | $3.4^{+2.1}_{-0.8}$ | ⋯ | ⋯ |
| | Median age (Gyr) (simulation) | $2.1^{+3.7}_{-1.6}$ | $2.3^{+3.8}_{-1.7}$ | $2.3^{+3.7}_{-1.8}$ | $2.7^{+3.6}_{-2.1}$ | $2.3^{+3.8}_{-1.8}$ | $1.6^{+3.4}_{-1.2}$ | $2.2^{+3.8}_{-1.7}$ | ⋯ | |
| | K-S (A09 simulation) | 0.5 | 0.4 | 0.4 | 0.4 | 0.4 | 0.6 | ⋯ | ⋯ | ⋯ |
| | K-S (J10 simulation) | 0.4 | 0.3 | 0.4 | 0.5 | 0.4 | 0.5 | ⋯ | ⋯ | ⋯ |
| | K-S (S21 simulation) | 0.5 | 0.5 | 0.4 | 0.4 | 0.4 | **0.7** | ⋯ | ⋯ | ⋯ |
| T | $H$ (pc) | $187^{+170}_{-64}$ | $190^{+183}_{-66}$ | $197^{+187}_{-70}$ | $191^{+164}_{-67}$ | $203^{+191}_{-75}$ | $189^{+168}_{-65}$ | $193^{+178}_{-68}$ | 4 | 0.4 |
| | $\sigma_w$ (km s$^{-1}$) | $13.1^{+5.0}_{-2.5}$ | $13.2^{+5.3}_{-2.6}$ | $13.4^{+5.3}_{-2.7}$ | $13.3^{+4.8}_{-2.6}$ | $13.7^{+5.4}_{-2.8}$ | $13.2^{+4.9}_{-2.5}$ | $13.3^{+5.2}_{-2.6}$ | ⋯ | ⋯ |

Table 8
(Continued)

| SpT | Quantity | B97 | B01 | B03 | SM08 | M18 | P20 | Median[a] | $N_{obs}$[b] | $N_{thick}$[c] |
|---|---|---|---|---|---|---|---|---|---|---|
| | Age (Gyr) (A09) | $2.6^{+2.8}_{-1.0}$ | $2.7^{+3.0}_{-1.0}$ | $2.8^{+3.1}_{-1.1}$ | $2.7^{+2.7}_{-1.0}$ | $2.9^{+3.2}_{-1.2}$ | $2.6^{+2.8}_{-1.0}$ | $2.7^{+2.9}_{-1.0}$ | ⋯ | ⋯ |
| | Age (Gyr) (J10) | $2.0^{+3.0}_{-0.9}$ | $2.0^{+3.2}_{-1.0}$ | $2.2^{+3.4}_{-1.0}$ | $2.1^{+2.9}_{-1.0}$ | $2.3^{+3.5}_{-1.1}$ | $2.0^{+3.0}_{-1.0}$ | $2.1^{+3.2}_{-1.0}$ | ⋯ | ⋯ |
| | Age (Gyr) (S21) | $3.3^{+3.7}_{-1.3}$ | $3.4^{+4.0}_{-1.3}$ | $3.5^{+4.1}_{-1.4}$ | $3.4^{+3.6}_{-1.4}$ | $3.7^{+4.2}_{-1.5}$ | $3.4^{+3.7}_{-1.3}$ | $3.4^{+3.9}_{-1.4}$ | ⋯ | ⋯ |
| | Median age (Gyr) (simulation) | $3.6^{+3.0}_{-2.4}$ | $3.5^{+3.0}_{-2.4}$ | $3.7^{+3.0}_{-2.5}$ | $3.6^{+3.0}_{-2.4}$ | $3.6^{+3.0}_{-2.5}$ | $3.7^{+3.0}_{-2.6}$ | $3.6^{+3.0}_{-2.5}$ | ⋯ | ⋯ |
| | K-S (A09 simulation) | 0.2 | 0.2 | 0.2 | 0.2 | 0.2 | 0.2 | ⋯ | ⋯ | ⋯ |
| | K-S (J10 simulation) | 0.3 | 0.3 | 0.2 | 0.3 | 0.2 | 0.3 | ⋯ | ⋯ | ⋯ |
| | K-S (S21 simulation) | 0.2 | 0.2 | 0.2 | 0.2 | 0.2 | 0.2 | ⋯ | ⋯ | ⋯ |

**Notes.** Model notation—B97: Burrows et al. (1997); B01: Burrows et al. (2001); B03: Baraffe et al. (2003); SM08: Saumon & Marley (2008); M18: Marley et al. (2018); P20: Phillips et al. (2020).
[a] Median values computed by combining all samples obtained from all models, not by standard error propagation formula. For M7–M9 dwarfs, only B97, B01, and B03 models are used; for L0–L4 dwarfs, the P20 models are excluded.
[b] The total number of UCDs in the magnitude-limited WISP and 3D-HST surveys analyzed in this study; simulations are compared to $N_{thin} = N_{obs} - N_{thick}$.
[c] Estimated number of thick disk contaminants; see Sections 3.4 and 4.2.
[d] Estimated probability that the age distribution of the simulated sample and that inferred from our measured number counts are distinct, as assessed through the K-S test. Values of K-S $\geqslant 0.7$ (highlighted in bold) are considered significantly distinct.

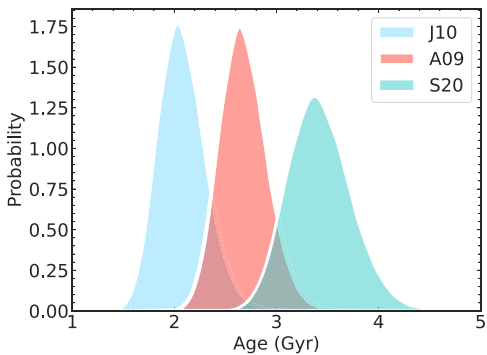

**Figure 13.** Comparison between the inferred age distributions for M7–M9 dwarfs in our sample based on the AVRs of J10, Aumer & Binney (2009, A09), and Sharma et al. (2021, S21). The J10 relations tend to predict the youngest ages while the S21 relations tend to predict the oldest ages. These distributions are based on the Baraffe et al. (2003) evolutionary models.

dwarfs, this sample nevertheless provides relatively weak constraints on the scaleheights and ages of the coldest brown dwarfs due to sensitivity and sample size limits. It is therefore useful to explore how deeper surveys planned for the recently launched JWST will improve upon these UCD population measurements. Ryan & Reid (2016) and Holwerda et al. (2018) have previously explored UCD source counts and identification in deep JWST imaging surveys; here we explore the expected yield from deep grism spectroscopic surveys, specifically PASSAGE; JWST Cycle 1 GO-1571, PI: Malkan, Malkan et al. 2021). This survey aims to study star formation across cosmic time by obtaining slitless grism spectra and imaging data in the F115W (0.9–1.3 μm), F150W (1.3–1.7 μm), and F200W (1.7–2.2 μm) passbands using the Near Infrared Imager and Slitless Spectrograph (NIRISS; Doyon et al. 2012; Willott et al. 2022) over various pointings at high Galactic latitudes ($|b| > 20^o$), with a goal of observing a total area of 0.16 deg$^2$. By increasing the depths to $J = 27$ (AB), this survey will produce a substantially deeper UCD sample. Several proposed programs with JWST are expected to reach comparable depths (Robertson 2021).

To simulate the expected number counts of UCDs in this survey, we followed a similar procedure as outlined in Section 3. The PASSAGE survey pointings are yet to be determined, so we chose a set of 124 random pointing at Galactic latitudes $|b| > 20^o$. We computed absolute magnitude/spectral type relations in the JWST/NIRISS filters using the methods described in Paper I (see Table 9), and assume a limiting magnitude of 27 (AB) in all three filters. We adopted the simulation parameters for the thin and thick disk population detailed in Table 4, using two vertical thin disk scaleheights (200 and 400 pc) that grid our results, and a common mass function between the thin disk, thick disk, and halo populations.

For the halo population, we assumed a flattened spheroid density distribution

$$\rho_{halo} = \left( \frac{R_\odot}{(r^2 + (z/q)^2)^{\frac{1}{2}}} \right)^n,$$

with parameters $q = 0.64$ and $n = 2.77$, a halo/thin disk density ratio of 0.25% (Jurić et al. 2008), and a uniform age distribution of 8–10 Gyr. Halo stars are typically older than 10 Gyr (Jofré & Weiss 2011; Guo et al. 2019); however, models do not cover this parameter range. Additionally, while both thick disk and halo stars and brown dwarfs are metal depleted compared to the thin disk, there is a lack of publicly available metal-poor evolutionary models for UCDs that span the late-M, L, T, and Y dwarf temperature range, so we deployed the solar-metallicity models of B01 and B03 to evolve our simulated sources. We follow the same procedure outlined in Section 3 by accounting for the binary fraction of UCDs, and the intrinsic scatter in the spectral type temperature and absolute magnitude-spectral type relations. We do not account for additional selection effects; however, we require that all sources should be detectable in F115W and F150W filters down to a magnitude limit of 27 (AB).

Figure 14 and Table 10 summarize the results of our simulation, showing the surface density and maximum distances of UCDs in the full PASSAGE survey as a function of spectral type. The sensitivity of JWST/NIRISS will allow us to detect late-M dwarfs to a limiting distance of ∼40 kpc, L dwarfs out to 10–30 kpc and T dwarfs out to 1–10 kpc, offering a tenfold increase in the limiting distances compared to deep UCD samples observed with HST (Table 1). However, the total

Table 9
Absolute Magnitude/Spectral Type Relations for M5–Y1 UCDs in NIRISS Passbands

| x | y | rms | $c_6$ | $c_5$ | $c_4$ | $c_3$ | $c_2$ | $c_1$ | $c_0$ |
|---|---|---|---|---|---|---|---|---|---|
| | | | | | | Coefficients | | | |
| SpT | $M_{F115W}$ | 0.39 | 4.29713588E-07 | −5.04229648E-05 | 2.22068501E-03 | −4.37392153E-02 | 3.02230231E-01 | 1.75203666E+00 | −1.70701962E+01 |
| SpT | $M_{F150W}$ | 0.40 | 3.81506552E-07 | −4.80917601E-05 | 2.37805487E-03 | −5.74753589E-02 | 6.67409384E-01 | −2.38264132E+00 | −1.22005576E+00 |
| SpT | $M_{F200W}$ | 0.40 | 7.59609823E-07 | −1.00377706E-04 | 5.31740074E-03 | −1.43601133E-01 | 2.05416628E+00 | −1.39462050E+01 | 3.72959094E+01 |

**Note.** Absolute magnitudes are computed as $M = \sum_{n=1}^{m} c_n \mathrm{SpT}^n$, with numerical spectral types mapped as M0: SpT = 10, L0: SpT = 20, T0: SpT = 30. These relations are valid for spectral types M5–Y1.

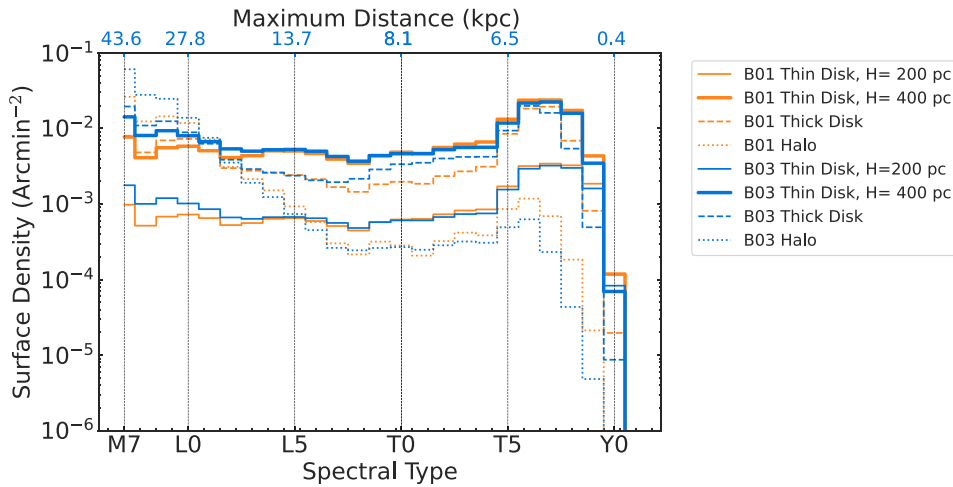

**Figure 14.** Surface density predictions in the PASSAGE survey for UCDs as a function of spectral type for thin disk (solid lines), thick disk (dashed lines), and halo populations (dotted lines). We model two thin disk populations with $H = 200$ pc (thin solid lines) and $H = 400$ pc (thick solid lines). We show simulations based on the Burrows et al. (2001, B01; blue lines) and Baraffe et al. (2003, B03; orange lines) evolutionary models.

Table 10
Expected Surface Densities (arcmin$^{-2}$) of M7–Y1 UCDs in the PASSAGE Survey

| Spectral Type | Thin Disk | | | | Thick Disk | | Halo | |
| --- | --- | --- | --- | --- | --- | --- | --- | --- |
| | Model = B01 | | Model = B03 | | Model = B01 | Model = B03 | Model = B01 | Model = B03 |
| | H = 200 pc | H = 400 pc | H = 200 pc | H = 400 pc | | | | |
| M7–M9 | 0.00217 | 0.0174 | 0.00397 | 0.0316 | 0.0198 | 0.0428 | 0.0531 | 0.113 |
| L | 0.00588 | 0.0459 | 0.00677 | 0.0528 | 0.0302 | 0.0359 | 0.0280 | 0.0299 |
| T | 0.0171 | 0.110 | 0.0156 | 0.100 | 0.0658 | 0.0704 | 0.00454 | 0.00283 |
| Y0–Y1 | <0.001 | <0.001 | <0.001 | <0.001 | <0.001 | <0.001 | <0.001 | <0.001 |
| Total | 0.0252 | 0.173 | 0.0265 | 0.185 | 0.116 | 0.149 | 0.0856 | 0.146 |

covered area in PASSAGES is about 4 times smaller than that of WISP and 3D-HST, resulting in a comparably sized sample as that examined here.

While not significantly improving overall sample statistics, PASSAGES will substantially improve our assessment of the thick disk and halo UCD population, with surface densities of thin disk, thick disk, and halo objects being roughly constant from late-M dwarfs to early T dwarfs. The higher proportion of thick disk and halo stars is due to the fact that JWST will easily reach the effective vertical edge of the thin disk. In addition, the longer cooling times for old thick disk and halo populations will boost their numbers among the late-L and T dwarfs. The exceptional distances probed result in a strong dependence of thin disk UCD surface densities as a function of vertical scaleheight, with a factor of 10 difference in L and T dwarf surface densities between $H = 200$ and $400$ pc. We also find that thick disk UCDs will outnumber thin disk UCDs down to mid-T, and halo UCDs will outnumber thin disk UCDs among late-M and early L dwarfs, depending on the thin disk scaleheight. Beyond spectral type T7, there is a significant drop-off in surface densities as the extreme faintness of the coldest brown dwarfs restricts their detection with a vertical scaleheight or less. Our simulations predict <1 Y dwarf in PASSAGE data, although the LF of these objects remains highly unconstrained (Kirkpatrick et al. 2021). We also note a small variation in predicted number counts among the late-M dwarfs between the B01 and B03 models, mirroring the discrepancies seen in our HST sample, and indicating an empirical path to explicitly testing the models.

Overall, our simulation predicts at least one thin disk, thick disk, or halo UCD per NIRISS imaging field (4.84 arcmin$^2$), with (depending on evolutionary model and scaleheight assumptions) 15–100 thin disk UCDs (cumulative surface densities 0.03–0.18 arcmin$^{-2}$), 70–90 thick disk UCDs (cumulative surface densities 0.12–0.15 arcmin$^{-2}$), and 50–90 halo UCDs (cumulative surface densities of 0.09–0.15 arcmin$^{-2}$), all relatively evenly distributed between late-M, L, and (thin/thick disk) T dwarfs. For comparison, Ryan & Reid (2016) predict surface densities of 0.02–0.1 arcmin$^{-2}$ for thin and thick disk M8–T5 dwarfs in JWST imaging down to $J = 27$ (AB), with thick disk objects dominating the sample. Our predictions indicate that UCDs in the PASSAGE fields will also be dominated by metal-poor thick disk and halo objects, so an appropriate simulation needs to take into account metallicity effects in both evolution and spectral energy distributions. A more complete study of these effects will be considered in a future study.

## 6. Summary

We summarize our findings as follows:

1. We attempted to reproduce the observed number counts in a deep spectral sample of UCDs from the WISP and 3D-HST HST/WFC3 surveys with Monte Carlo population simulations that combine assumptions for the star formation history, mass function, local LF, and spatial model of the Galactic UCD population.

2. By comparing simulations that varied the thin disk vertical scaleheight and choice of evolutionary model, we

inferred the scaleheight distribution of UCDs as a function of spectral subclass, a proxy of population age, and hence the star formation and evolutionary history of our mixed stellar and substellar sample.

3. We found a late-M dwarf scaleheight of $249^{+48}_{-61}$ pc, a value smaller than but consistent with prior deep HST imaging samples and ground-based survey measurements. We also found an L dwarf scaleheight of $153^{+56}_{-30}$ pc and a T dwarf scaleheight of $175^{+149}_{-56}$ pc, both considerably smaller than prior space-based and ground-based deep imaging surveys.

4. Using transformations between scaleheight, velocity dispersion, and age, we determined population ages of $3.6^{+0.8}_{-1.0}$ Gyr for late-M dwarfs, $2.1^{+0.9}_{-0.5}$ Gyr for L dwarfs, and $2.4^{+2.4}_{-0.8}$ Gyr for T dwarfs. While there is some variance between these spatially based ages and velocity dispersion-based ages measured in the local UCD population, and systematic effects in age-velocity-scaleheight transformations contribute significantly to the uncertainties in the absolute ages (1–2 Gyr), the relative drop-off in age measured between late-M dwarfs and L and T dwarfs is consistent with predictions based on brown dwarf population simulations.

5. We used our simulations to predict the expected UCD yield in the deep JWST PASSAGE spectral survey, which will reach distances of ∼40 kpc for late-M dwarfs, 10–30 kpc for L dwarfs, and 1–10 kpc for T dwarfs. The smaller area of PASSAGES compared to WISP and 3D-HST results in a comparably sized spectral sample of 135-280 UCDs, but dominated by thick disk and halo objects. Thus, metallicity effects in evolution and observable properties will be more important in JWST surveys than HST equivalents.

This series provides a first glimpse into the utility and limitations of using deep spectral samples of UCDs for investigations of the Galaxy at large. While the HST sample examined here greatly expands upon prior deep spectral surveys, and has far greater fidelity than comparable image-based surveys, sensitivity limits nevertheless restrict both the sample size (particularly for late-L, T, and Y dwarfs) and accessibility to major Galactic populations (halo and thick disk subdwarfs). We also find important systematic differences between current evolutionary models, particularly for late-M and early L dwarfs, which sample the largest distances in our survey. Hence, while our determinations for the scaleheights and ages of UCDs align with prior deep imaging and local kinematic studies, our uncertainties remain sufficiently large to limit our ability to critically assess evolution-induced age variations and explore star formation parameters in detail. Fortunately, larger and deeper spectral surveys are planned in the forthcoming space missions JWST, SPHEREx (Doré et al. 2014), the Nancy Grace Roman Space Telescope (Spergel et al. 2015), and the Euclid telescope (Laureijs et al. 2011; Solano et al. 2021). Our analysis of the JWST PASSAGE program predicts a UCD spectral sample extending to tens of kiloparsecs in distance, with a majority of thick disk and halo sources. For comparison, Solano et al. (2021) have predicted millions of UCDs in the Euclid wide-field survey in multiple imaging filters, while the ground-based Vera Rubin Observatory's Legacy Survey of Space and Time (LSST; Ivezić et al. 2019) is expected to detect $>10^6$ UCDs with multi-epoch and multicolor photometry and astrometry. With these near-term improvements in sample size and fidelity, the full potential of UCD tracers for Galactic archeology studies will be realized.

## 7. Absolute Magnitude/Spectral Type Relations for JWST/NIRISS filters

Table 9 lists absolute magnitude/spectral type relations for M5–Y1 UCDs in JWST/NIRISS filters. The derivation of absolute magnitude relations follows the methodology in Paper I. Magnitudes were measured by convolving NIRISS filter profiles with low-resolution spectra of M5–T9 UCDs from the SpeX Prism Library (Burgasser 2014). Additional WFC3 spectra of Y0–Y1 were obtained from Schneider et al. (2015). We computed color corrections between magnitudes in the NIRISS filters and magnitudes in either the MKO J of Two Micron All Sky Survey H filters, then applied these corrections to the absolute magnitude relations of Dupuy & Liu (2012) for spectral types earlier than T8 and Kirkpatrick et al. (2021) for spectral types T8–Y2. We propagated uncertainties by random sampling and derived polynomial coefficients by fitting a sixth-degree polynomial, clipping $3\sigma$ outliers.

## 8. HST/WFC3 Pointings

Table 3 provides a list of the pointings covered by the WISP and 3D-HST surveys, grism integration times, and limiting imaging magnitudes in the filters used.

This research is based on observations made with the NASA/ESA Hubble Space Telescope obtained from the Space Telescope Science Institute, which is operated by the Association of Universities for Research in Astronomy, Inc., under NASA contract No. NAS 5-26555. These observations are associated with programs GO-12177 and GO-12328 (3D-HST), and programs GO-11696, GO-12283, GO-12568, GO-12902, GO-13352, GO-13517, and GO-14178 (WISP). Support for this work was provided by NASA through the NASA Hubble Fellowship grant No. HST-HF2-51447.001-A awarded by the Space Telescope Science Institute to NASA, under contract No. NAS5-26555. This research has benefited from the SpeX Prism Library, maintained by Adam Burgasser at http://www.browndwarfs.org/spexprism. C.A. thanks Lucianne Walkowicz, Adam Miller, Ivelina Momcheva, and members of the WISP team for help in analyzing survey data; and the LSSTC Data Science Fellowship Program, which is funded by LSSTC, NSF Cybertraining Grant #1829740, the Brinson Foundation, and the Moore Foundation. C.A. acknowledges funding from the UC Office of the President UC-HBCU Pathways Program. This material is based upon work supported by NASA under grant No. NNX16AF47G issued through the Astrophysics Data Analysis Program. Portions of this work were conducted at the University of California San Diego, which was built on the unceded territory of the Kumeyaay Nation, whose people continue to maintain their political sovereignty and cultural traditions as vital members of the San Diego community.

*Software:* Astropy (Price-Whelan et al. 2018), SPLAT (Burgasser & Splat Development Team 2017), SciPy (Virtanen et al. 2020), Matplotlib (Hunter 2007), Seaborn (Waskom et al. 2014), NumPy (Harris et al. 2020), Pandas (McKinney 2010), Axe (Kümmel et al. 2009).

## ORCID iDs

Christian Aganze https://orcid.org/0000-0003-2094-9128
Adam J. Burgasser https://orcid.org/0000-0002-6523-9536
Mathew Malkan https://orcid.org/0000-0001-6919-1237
Christopher A. Theissen https://orcid.org/0000-0002-9807-5435
Roberto A. Tejada Arevalo https://orcid.org/0000-0001-6708-3427
Chih-Chun Hsu https://orcid.org/0000-0002-5370-7494
Daniella C. Bardalez Gagliuffi https://orcid.org/0000-0001-8170-7072
Russell E. Ryan, Jr. https://orcid.org/0000-0003-0894-1588
Benne Holwerda https://orcid.org/0000-0002-4884-6756

## References

Aganze, C., Burgasser, A. J., Faherty, J. K., et al. 2016, AJ, 151, 46
Aganze, C., Burgasser, A. J., Malkan, M., et al. 2022, ApJ, 924, 114
Aihara, H., Arimoto, N., Armstrong, R., et al. 2018, PASJ, 70, S4
Antoja, T., Helmi, A., Romero-Gómez, M., et al. 2018, Natur, 561, 360
Atek, H., Malkan, M., McCarthy, P., et al. 2010, ApJ, 723, 104
Aumer, M., Binney, J., & Schönrich, R. 2016, MNRAS, 462, 1697
Aumer, M., & Binney, J. J. 2009, MNRAS, 397, 1286
Bahcall, J., & Soneira, R. 1981, ApJS, 47, 357
Baraffe, I., Chabrier, G., Barman, T., Allard, F., & Hauschildt, P. 2003, A&A, 402, 701
Bardalez Gagliuffi, D. C., Burgasser, A. J., Schmidt, S. J., et al. 2019, ApJ, 883, 205
Basri, G., & Reiners, A. 2006, AJ, 132, 663
Bastian, N., Covey, K. R., & Meyer, M. R. 2010, ARA&A, 48, 339
Belokurov, V., Erkal, D., Evans, N. W., Koposov, S. E., & Deason, A. J. 2018, MNRAS, 478, 611
Bensby, T., Feltzing, S., & Lundström, I. 2003, A&A, 410, 527
Bertelli, G., Girardi, L., Marigo, P., & Nasi, E. 2008, A&A, 484, 815
Bird, J. C., Kazantzidis, S., Weinberg, D. H., et al. 2013, ApJ, 773, 43
Blake, C. H., Charbonneau, D., & White, R. J. 2010, ApJ, 723, 684
Bland-Hawthorn, J., & Gerhard, O. 2016, ARA&A, 54, 529
Bochanski, J., Hawley, S., Covey, K., et al. 2010, AJ, 139, 2679
Borucki, W. J., Koch, D., Basri, G., et al. 2010, Sci, 327, 977
Boubert, D., Guillochon, J., Hawkins, K., et al. 2018, MNRAS, 479, 2789
Bovy, J. 2017, MNRAS, 470, 1360
Bovy, J., & Rix, H.-W. 2013, ApJ, 779, 115
Bovy, J., Rix, H.-W., & Hogg, D. W. 2012, ApJ, 751, 131
Brammer, G., van Dokkum, P., Franx, M., et al. 2012, ApJS, 200, 13
Buck, T., Obreja, A., Macciò, A. V., et al. 2020, MNRAS, 491, 3461
Burgasser, A. 2004, ApJS, 155, 191
Burgasser, A., Kirkpatrick, J., Burrows, A., et al. 2003, ApJ, 592, 1186
Burgasser, A. J. 2007, ApJ, 659, 655
Burgasser, A. J. 2014, ASI Conf. Ser. 11, Int. Workshop on Stellar Spectral Libraries, ed. H. P. Singh, P. Prugniel, & I. Vauglin, (Bangalore: ASI), 7
Burgasser, A. J., Kirkpatrick, J. D., Cruz, K. L., et al. 2006, ApJS, 166, 585
Burgasser, A. J., Logsdon, S. E., Gagné, J., et al. 2015, ApJS, 220, 18
Burgasser, A. J., Marley, M. S., Ackerman, A. S., et al. 2002, ApJL, 571, L151
Burgasser, A. J. & Splat Development Team 2017, ASI Conf. Ser. 14, Int. Workshop on Spectral Stellar Libraries (Bangalore: ASI), 7
Burningham, B., Cardoso, C., Smith, L., et al. 2013, MNRAS, 433, 457
Burrows, A., Hubbard, W. B., Lunine, J. I., & Liebert, J. 2001, RvMP, 73, 719
Burrows, A., Marley, M., Hubbard, W. B., et al. 1997, ApJ, 491, 856
Caiazzo, I., Heyl, J. S., Richer, H., & Kalirai, J. 2017, arXiv:1702.00091
Carnero Rosell, A., Santiago, B., dal Ponte, M., et al. 2019, MNRAS, 489, 5301
Chabrier, G., & Mera, D. 1997, A&A, 328, 83
Chen, B., Stoughton, C., Smith, J. A., et al. 2001, ApJ, 553, 184
Cruz, K., Reid, I., Kirkpatrick, J., et al. 2007, AJ, 133, 439
Cruz, K. L., Reid, I. N., Liebert, J., Kirkpatrick, J. D., & Lowrance, P. J. 2003, AJ, 126, 2421
Cushing, M., Rayner, J., & Vacca, W. 2005, ApJ, 623, 1115
De Silva, G. M., Freeman, K. C., Bland-Hawthorn, J., et al. 2015, MNRAS, 449, 2604
de Vaucouleurs, G., & Pence, W. D. 1978, AJ, 83, 1163
Doré, O., Bock, J., Ashby, M., et al. 2014, arXiv:1412.4872
Doyon, R., Hutchings, J. B., Beaulieu, M., et al. 2012, Proc. SPIE, 8442, 84422R
Dupuy, T., & Liu, M. 2012, ApJS, 201, 19
Erkal, D., & Belokurov, V. A. 2020, MNRAS, 495, 2554
Fontanive, C., Biller, B., Bonavita, M., & Allers, K. 2018, MNRAS, 479, 2702
Fouesneau, M., Rix, H.-W., von Hippel, T., Hogg, D. W., & Tian, H. 2019, ApJ, 870, 9
Freeman, K. C. 1987, ARA&A, 25, 603
Gerasimov, R., Burgasser, A. J., Homeier, D., et al. 2022, ApJ, 930, 24
Giavalisco, M., Ferguson, H. C., Koekemoer, A. M., et al. 2004, ApJ, 600, L93
Gillessen, S., Eisenhauer, F., Trippe, S., et al. 2009, ApJ, 692, 1075
Gilmore, G., Wyse, R. F. G., & Jones, J. B. 1995, AJ, 109, 1095
Gould, A., Bahcall, J., & Flynn, C. 1997, ApJ, 482, 913
Guo, J.-C., Zhang, H.-W., Huang, Y., et al. 2019, RAA, 19, 008
Harris, C. R., Millman, K. J., van der Walt, S. J., et al. 2020, Natur, 585, 357
Hawkins, K., Jofré, P., Masseron, T., & Gilmore, G. 2015, MNRAS, 453, 758
Hayden, M. R., Bovy, J., Holtzman, J. A., et al. 2015, ApJ, 808, 132
Haywood, M., Di Matteo, P., Lehnert, M. D., Katz, D., & Gómez, A. 2013, A&A, 560, A109
Helmi, A., Babusiaux, C., Koppelman, H. H., et al. 2018, Natur, 563, 85
Helmi, A., White, S. D. M., de Zeeuw, P. T., & Zhao, H. 1999, Natur, 402, 53
Holmberg, J., Nordström, B., & Andersen, J. 2007, A&A, 475, 519
Holwerda, B. W., Bridge, J. S., Ryan, R. J., et al. 2018, A&A, 620, A132
Holwerda, B. W., Trenti, M., Clarkson, W., et al. 2014, ApJ, 788, 77
Hopkins, P. F., Hernquist, L., Cox, T. J., Younger, J. D., & Besla, G. 2008, ApJ, 688, 757
Howell, S. B., Sobeck, C., Haas, M., et al. 2014, PASP, 126, 398
Hsu, C.-C., Burgasser, A. J., Theissen, C. A., et al. 2021, ApJS, 257, 45
Hunter, J. D. 2007, CSE, 9, 90
Ivezić, Ž., Beers, T. C., & Jurić, M. 2012, ARA&A, 50, 251
Ivezić, Ž., Kahn, S. M., Tyson, J. A., et al. 2019, ApJ, 873, 111
Jahreiß, H., & Wielen, R. 1997, in ESA Special Publication, 402, Hipparcos—Venice '97, ed. B. Battrick, M. A. C. Perryman, & P. L. Bernacca, 675
Jofré, P., & Weiss, A. 2011, A&A, 533, A59
Jurić, M., Ivezić, Ž., Brooks, A., et al. 2008, ApJ, 673, 864
Just, A., & Jahreiß, H. 2010, MNRAS, 402, 461
Kakazu, Y., Hu, E. M., Liu, M. C., et al. 2010, ApJ, 723, 184
Kerins, E. 1997, A&A, 328, 5
Kilic, M., Munn, J. A., Harris, H. C., et al. 2017, ApJ, 837, 162
Kirkpatrick, J. 2005, ARA&A, 43, 195
Kirkpatrick, J., Schneider, A., Fajardo-Acosta, S., et al. 2014, ApJ, 783, 122
Kirkpatrick, J. D., Gelino, C. R., Cushing, M. C., et al. 2012, ApJ, 753, 156
Kirkpatrick, J. D., Gelino, C. R., Faherty, J. K., et al. 2021, ApJS, 253, 7
Kirkpatrick, J. D., Martin, E. C., Smart, R. L., et al. 2019, ApJS, 240, 19
Koppelman, H. H., Helmi, A., Massari, D., Price-Whelan, A. M., & Starkenburg, T. K. 2019, A&A, 631, L9
Kümmel, M., Walsh, J. R., Pirzkal, N., Kuntschner, H., & Pasquali, A. 2009, PASP, 121, 59
Lacey, C. G. 1984, MNRAS, 208, 687
Laporte, C. F. P., Minchev, I., Johnston, K. V., & Gómez, F. A. 2019, MNRAS, 485, 3134
Laureijs, R., Amiaux, J., Arduini, S., et al. 2011, arXiv:1110.3193
Leggett, S., Ruiz, M., & Bergeron, P. 1998, ApJ, 497, 294
Lépine, S., & Scholz, R.-D. 2008, ApJL, 681, L33
Liu, M. C., Leggett, S. K., Golimowski, D. A., et al. 2006, ApJ, 647, 1393
Looper, D. L., Gelino, C. R., Burgasser, A. J., & Kirkpatrick, J. D. 2008, ApJ, 685, 1183
Ma, X., Hopkins, P. F., Wetzel, A. R., et al. 2017, MNRAS, 467, 2430
Mackereth, J. T., Bovy, J., Leung, H. W., et al. 2019, MNRAS, 489, 176
Mackereth, J. T., Bovy, J., Schiavon, R. P., et al. 2017, MNRAS, 471, 3057
Malhan, K., Ibata, R. A., & Martin, N. F. 2018, MNRAS, 481, 3442
Malkan, M. A., Alavi, A., Atek, H., et al. 2021, PASSAGE-Parallel Application of Slitless Spectroscopy to Analyze Galaxy Evolution, JWST Proposal ID 1571, STSci
Manjavacas, E., Apai, D., Zhou, Y., et al. 2019, AJ, 157, 101
Marley, M., Saumon, D., Morley, C., & Fortney, J. 2018, Sonora 2018: Cloud-free, solar composition, solar C/O substellar atmosphere models and spectra, nc_m+0.0_co1.0_v1.0, Zenodo, doi:10.5281/zenodo.1309035
Martig, M., Minchev, I., & Flynn, C. 2014, MNRAS, 443, 2452
Martín, E. L., Lodieu, N., Pavlenko, Y., & Béjar, V. J. S. 2018, ApJ, 856, 40
Masters, D., McCarthy, P., Burgasser, A. J., et al. 2012, ApJL, 752, L14
McKinney, W. 2010, in Proc. 9th Python in Science Conf. 56, ed. S. van der Walt & J. Millman (Austin, TX: SciPy), 56, https://pandas.pydata.org/about/citing.html
Metchev, S., Kirkpatrick, J., Berriman, G., & Looper, D. 2008, ApJ, 676, 1281
Minchev, I., Martig, M., Streich, D., et al. 2015, ApJL, 804, L9

Momcheva, I. G., Brammer, G. B., van Dokkum, P. G., et al. 2016, ApJS, 225, 27
Myeong, G. C., Evans, N. W., Belokurov, V., Sanders, J. L., & Koposov, S. E. 2018, ApJL, 856, L26
Naidu, R. P., Conroy, C., Bonaca, A., et al. 2020, ApJ, 901, 48
Nordström, B., Mayor, M., Andersen, J., et al. 2004, A&A, 418, 989
Parzen, E. 1962, Ann. Math. Statist, 33, 1065
Pecaut, M. J., & Mamajek, E. E. 2013, ApJS, 208, 9
Phillips, M. W., Tremblin, P., Baraffe, I., et al. 2020, A&A, 637, A38
Pirzkal, N., Burgasser, A., Malhotra, S., et al. 2009, ApJ, 695, 1591
Pirzkal, N., Sahu, K., Burgasser, A., et al. 2005, ApJ, 622, 319
Pirzkal, N., Xu, C., Malhotra, S., et al. 2004, ApJS, 154, 501
Price-Whelan, A. M., Johnston, K. V., Sheffield, A. A., Laporte, C. F. P., & Sesar, B. 2015, MNRAS, 452, 676
Price-Whelan, A. M., Sipőcz, B. M., Günther, H. M., et al. 2018, AJ, 156, 123
Reid, I. 2003, AJ, 126, 2449
Reyle, C., Delorme, P., Willott, C. J., et al. 2010, A&A, 522, A112
Reylé, C., Jardine, K., Fouqué, P., et al. 2021, A&A, 650, A201
Ricker, G. R., Winn, J. N., Vanderspek, R., et al. 2014, Proc. SPIE, 9143, 914320
Robertson, B. E. 2021, arXiv:2110.13160
Rojas-Ayala, B., Covey, K. R., Muirhead, P. S., & Lloyd, J. P. 2012, ApJ, 748, 93
Ryan, R. E. J., Hathi, N., Cohen, S., & Windhorst, R. 2005, ApJL, 631, L159
Ryan, R. E. J., & Reid, I. N. 2016, AJ, 151, 92
Ryan, R. E. J., Thorman, P. A., Schmidt, S. J., et al. 2017, ApJ, 847, 53
Ryan, R. E. J., Thorman, P. A., Yan, H., et al. 2011, ApJ, 739, 83
Sanders, J. L., & Das, P. 2018, MNRAS, 481, 4093
Saumon, D., & Marley, M. 2008, ApJ, 689, 1327
Schneider, A. C., Burgasser, A. J., Gerasimov, R., et al. 2020, ApJ, 898, 77
Schneider, A. C., Cushing, M. C., Kirkpatrick, J. D., et al. 2015, ApJ, 804, 92
Sellwood, J. A., & Binney, J. J. 2002, MNRAS, 336, 785
Sharma, S., Hayden, M. R., Bland-Hawthorn, J., et al. 2021, MNRAS, 506, 1761
Sharma, S., Stello, D., Bland-Hawthorn, J., et al. 2019, MNRAS, 490, 5335
Sharma, S., Stello, D., Buder, S., et al. 2018, MNRAS, 473, 2004
Sheinis, A., Anguiano, B., Asplund, M., et al. 2015, JATIS, 1, 035002
Shipp, N., Drlica-Wagner, A., Balbinot, E., et al. 2018, ApJ, 862, 114
Skelton, R. E., Whitaker, K. E., Momcheva, I. G., et al. 2014, ApJS, 214, 24
Skrzypek, N., Warren, S. J., & Faherty, J. K. 2016, A&A, 589, A49
Solano, E., Gálvez-Ortiz, M. C., Martín, E. L., et al. 2021, MNRAS, 501, 281
Sorahana, S., Nakajima, T., & Matsuoka, Y. 2019, ApJ, 870, 118
Spergel, D., Gehrels, N., Baltay, C., et al. 2015, arXiv:1503.03757
Spitzer, L. J., & Schwarzschild, M. 1953, ApJ, 118, 106
Stauffer, J., Schultz, G., & Kirkpatrick, J. 1998, ApJL, 499, L199
The Dark Energy Survey Collaboration 2005, arXiv:astro-ph/0510346
Tolstoy, E., Hill, V., & Tosi, M. 2009, ARA&A, 47, 371
Toth, G., & Ostriker, J. P. 1992, ApJ, 389, 5
van der Kruit, P. C. 1988, A&A, 192, 117
van Leeuwen, F. 2007, A&A, 474, 653
van Vledder, I., van der Vlugt, D., Holwerda, B. W., et al. 2016, MNRAS, 458, 425
Veyette, M. J., Muirhead, P. S., Mann, A. W., et al. 2017, ApJ, 851, 26
Virtanen, P., Gommers, R., et al. 2020, NatMe, 17, 261
Warren, S. J., Ahmed, S., & Laithwaite, R. C. 2021, OJAp, 4, 4
Waskom, M., Botvinnik, O., Hobson, P., et al. 2014, seaborn: v0.5.0 (November 2014), v0.5.0, Zenodo, doi:10.5281/zenodo.12710
Willott, C. J., Doyon, R., Albert, L., et al. 2022, PASP, 134, 025002
York, D. G., Adelman, J., Anderson, J. E., Jr., et al. 2000, AJ, 120, 1579
Yuan, Z., Myeong, G. C., Beers, T. C., et al. 2020, ApJ, 891, 39
Zapatero Osorio, M. R., Martín, E. L., Béjar, V. J. S., et al. 2007, ApJ, 666, 1205
Zhang, Z. H., Burgasser, A. J., Gálvez-Ortiz, M. C., et al. 2019, MNRAS, 486, 1260
Zhang, Z. H., Pinfield, D. J., Gálvez-Ortiz, M. C., et al. 2017, MNRAS, 464, 3040
Zhao, G., Zhao, Y.-H., Chu, Y.-Q., Jing, Y.-P., & Deng, L.-C. 2012, RAA, 12, 723